\newif\ifarxiv\arxivtrue
\pdfoutput=1 
\ifarxiv
\documentclass[acmsmall,screen,nonacm]{acmart}
\else
\documentclass[acmsmall,screen]{acmart}
\fi
\usepackage{pervasives}

\ifarxiv
\setcopyright{none}
\else

\setcopyright{acmlicensed}
\acmPrice{}
\acmDOI{10.1145/3485488}
\acmYear{2021}
\copyrightyear{2021}
\acmSubmissionID{oopsla21main-p49-p}
\acmJournal{PACMPL}
\acmVolume{5}
\acmNumber{OOPSLA}
\acmArticle{111}
\acmMonth{10}


\fi

\begin{document}

\title{Solver-based Gradual Type Migration}

\author{Luna Phipps-Costin}
\affiliation{\city{Amherst}\institution{University of Massachusetts Amherst}\country{United States}}
\author{Carolyn Jane Anderson}
\affiliation{\city{Wellesley}\institution{Wellesley College}\country{United States}}
\author{Michael Greenberg}
\affiliation{\city{Los Angeles}\institution{Pomona College}\country{United States}}
\author{Arjun Guha}
\affiliation{\city{Boston}\institution{Northeastern University}\country{United States}}

\begin{abstract}

Gradually typed languages allow programmers to mix statically and dynamically 
typed code, enabling them to incrementally reap the benefits of static typing 
as they add type annotations to their code. However, this type migration 
process is typically a manual effort with limited tool support.
This paper examines the problem of \emph{automated type migration}: 
given a dynamic program, infer additional or improved type annotations.

Existing type migration algorithms prioritize different goals, such as 
maximizing type precision, maintaining compatibility with unmigrated code, 
and preserving the semantics of the original program. We argue that 
the type migration problem involves fundamental compromises: optimizing 
for a single goal often comes at the expense of others. Ideally, a type migration 
tool would flexibly accommodate a range of user priorities.

We present \ourtool, a new approach to automated type migration for
the gradually-typed lambda calculus with some extensions. Unlike prior work,
which relies on custom solvers, \ourtool produces constraints for
an off-the-shelf MaxSMT solver. This allows us to easily express objectives,
such as minimizing the number of necessary syntactic coercions, and constraining the
type of the migration to be compatible with unmigrated code.

We present the first comprehensive evaluation of GTLC type migration
algorithms, and compare \ourtool{} to four other tools from the literature. Our
evaluation uses prior benchmarks, and a new set of ``challenge problems.''
Moreover, we design a new evaluation methodology that highlights the subtleties
of gradual type migration. In addition, we apply \ourtool{} to a suite of
benchmarks for Grift, a programming language based on the GTLC. \ourtool{} is
able to reconstruct all human-written annotations on all but one program.
\end{abstract}

\maketitle

\section{Introduction}

Gradually typed languages allow programmers to freely mix statically and 
dynamically typed code. This enables users to add static types gradually,
providing the benefits of static typing without requiring the entirety of a
 codebase to be overhauled at once~\cite{th:migration,siek:gtlc}. Over the past decade, 
gradually typed dialects of several mainstream languages, such as JavaScript, 
Python, and Ruby, have become established in industry. However,
the process of \emph{migrating} an untyped program to use gradual types has largely
 remained a labor-intensive manual effort. Just as type inference facilitates static typing, 
 type migration tools have the potential to make gradual typing easier to use.

However, automating type migration is a challenging problem. Even if we consider a
small language, such as the \emph{gradually typed lambda calculus}
(GTLC)~\cite{siek:gtlc}, and limit ourselves to modifying existing type annotations, a
single program may have many possible migrations. Existing approaches either
produce a single migration~\cite{wright:soft-typing,flanagan:spidey,scheme-to-ml,
siek:gti,rastogi:gti}, or a menu of possible migrations without guidance on which to
 select~\cite{campora:migrating,migeed:decidable}. How should we
choose among the migrations produced by various approaches?

In this paper, we present a new approach to gradual type migration for the GTLC,
and the first comprehensive evaluation of prior work in this area. We first
illustrate the tension between
\emph{precise type migrations} that produce informative type annotations,
\emph{safe type migrations} that do not introduce new dynamic errors, and
\emph{compatible type migrations} that preserve interoperability with
other, unmigrated code. We show that prior
approaches navigate these tradeoffs in different ways. Some favor making types as
precise as possible, even if the increased precision compromises safety or
compatibility. Others favor compatibility and safety at the expense of
precision. Finally, many approaches  statically reject programs
that may have dynamic type errors.

We present \ourtool, a new approach to type migration that navigates these
tradeoffs as follows. (1)~\ourtool does not statically reject any closed
programs, including programs that obviously crash with a dynamic error.
(2)~\ourtool always produces safe type migrations, thus does not introduce
new dynamic errors. (3)~\ourtool is the first approach that allows the
user to chose between precise migrations and compatible migrations, and
we show that there are different situations in which the user may prefer
one or the other.

Under the hood, \ourtool differs from prior work in two key ways.
First, whereas prior work relies on custom constraint
solvers, \ourtool generates constraints for an off-the-shelf MaxSMT
solver~\cite{z3maxsmt}. This makes it easy to add constraints and language
features, as we demonstrate by extending the GTLC in several ways and supporting
the Grift gradually typed language~\cite{kuhlenschmidt:grift}.

Second, using a general-purpose solver is what allows \ourtool to readily support
multiple kinds of migrations. We accomplish this by
 using the MaxSMT solver in a two-stage process. We first formulate a 
 MaxSMT problem with an objective function that synthesizes precise types. 
 The reconstructed type of the program may not be compatible with all contexts, but it reveals the 
 (potentially higher-order) interface of the program. We then formulate 
 new constraints on the type of the program to enforce compatibility,
 and use the MaxSMT solver a second time to 
 produce a new solution.

Our evaluation compares \ourtool to four other type migration 
approaches using a two-part evaluation suite: a set of existing benchmarks by \citet{migeed:decidable}, and a new set of ``challenge problems'' that we devise. 
We also design an evaluation methodology that reflects the subtleties of type migration. 
Although different approaches to type migration prioritize different goals, \ourtool
performs as well or better than existing tools on all prior benchmark suites.
We design our ``challenge problems'' to highlight the strengths and weaknesses
of all approaches, including \ourtool.
 Finally, we apply \ourtool to a suite 
of Grift programs from \citet{kuhlenschmidt:grift}, and find that it reproduces all 
hand-written type annotations except in one case.

\paragraph{Limitations}

\ourtool{} focuses on type migration for the core GTLC, which merely extends
simple types with an unknown (\tdyn). This allows us to directly compare \ourtool{}
to  other GTLC type migration algorithms (\cref{sec:evaluation}). But,
it does limit the scope of our work and the conclusions we can draw:

\begin{enumerate}

\item This  paper does \emph{not}
consider union types, polymorphism, subtyping, recursive types, and other
features that are necessary to build more complete programming languages.

\item Our evaluation on the GTLC uses small, artificial programs. These
benchmarks illustrate tricky cases where different approaches produce different
results, but they do not represent the structure of real-world programs.
\ourtool{} has a frontend for the Grift programming language, which we use to
evaluate on the benchmarks presented by \citet{kuhlenschmidt:grift}, but these
are also small.

\item This paper focuses on type migration for the guarded semantics of gradual
types. Many gradually typed languages, including TypeScript, use
alternative semantics, which we briefly discuss in \cref{sec:related}.

\end{enumerate}

\paragraph{Contributions} Our key contributions are as follows:
\begin{enumerate}

  \item We illustrate the tension between the many goals of type migration,
  and argue that type migration tools should give users the ability to choose
  between different goals (\cref{sec:what_matters} and \cref{sec:type_migration}).

  \item We present the \ourtool{} approach to type migration, which formulates
  constraints for an off-the-shelf MaxSMT solver (\cref{sec:smt}). \ourtool{}
  supports the GTLC and additional language features required to support
  the Grift gradually typed language (\cref{sec:smt_extensions}).

  \item We present a new set of type migration ``challenge problems'' that
  illustrate the strengths and weaknesses of different approaches to type
  migration (\cref{sec:evaluation}).

  \item We present a comprehensive comparison of \numtools{} approaches to type
  migration (including ours), using a new evaluation methodology. For this
  comparison, we implement a unified framework for running, evaluating, and
  validating type migration algorithms.

  \item Finally, we contribute re-implementations of the type migration
  algorithms from \citet{campora:migrating} and \citet{rastogi:gti}. Ours is the 
  first publicly available implementation of \citet{rastogi:gti}.

\end{enumerate}

\paragraph{Artifact} The artifact for this paper is available
at \url{https://doi.org/10.5281/zenodo.5141479}.

\section{What Matters for Type Migration?}\label{sec:what_matters}
\lstset{language=ocaml}

When designing a type migration tool, we must consider several important
questions:

\begin{enumerate}

  \item A key goal of type migration is to improve the precision of type 
  annotations. However, there are often multiple ways to improve type precision~\cite{migeed:decidable} that induce different run-time checks. 
For any given type migration system, we must therefore ask the question, 
\emph{Can a user choose between several alternative migrations?}

  \item When the migrated code is only a fragment of a larger
  codebase, increasing type precision can 
  introduce type errors at the boundaries between migrated and unmigrated
  code~\cite{rastogi:gti}. Thus we must ask, \emph{Does the migrated code
  remain compatible with other, unmigrated code?}

  \item A type migration tool may also uncover potential run-time errors. However,
  these errors may be unreachable, or only occur in certain configurations or on certain platforms.
  Thus we must ask, \emph{Should a migration
  turn (potential) run-time errors into static type errors?}

  \item Finally, safe gradually typed languages introduce checks
  that enforce type safety at run-time. Making a type more precise 
  can alter these checks, affecting run-time behavior. Thus we must ask,
  \emph{Does the migrated program preserve the behavior of the original
  program?}

\end{enumerate}

This section explores these questions with examples from the
gradually-typed lambda calculus (GTLC) with some modest extensions. We write
programs in an
OCaml-like syntax with explicit type annotations. The type \tdyn is the
\emph{unknown type} (also known as the dynamic type or the any type), which is
compatible with all types. Under the hood, converting to and from the \tdyn type
introduces coercions~\cite{henglein:coercions}; these coercions can fail at run-time
with a dynamic type error.

\paragraph{Type migration can introduce new static errors}

\begin{wrapfigure}{r}{0.24\textwidth}
\begin{lstlisting}[aboveskip=-10pt]
let A (x : STAR) = 
  let _ = x + 10 in
  x ()
\end{lstlisting}
\caption{Reachable error.}
\label{ex-ts-reachable-crash}
\end{wrapfigure}

\Cref{ex-ts-reachable-crash} shows a function that uses its \tdyn-typed
argument first as a number and then as a function. Since \tdyn is compatible
with all types, the function is well-typed, but guaranteed to produce a dynamic
type error when applied. In this case, it seems harmless for a type
migration tool to turn this dynamic type error into a static type error.

\begin{wrapfigure}{r}{0.24\textwidth}
\vspace{-20pt}
\begin{lstlisting}
let B (x : STAR) =
  if false then 
    let _ = x + 10 in
    x ()
  else
    x ()
\end{lstlisting}
\caption{Unreachable error.}
\label{ex-ts-unreachable}
\vspace{-10pt}
\end{wrapfigure}
  
However, it is also possible for the crashing expression to be unreachable.
\Cref{ex-ts-unreachable} wraps the same dynamic error in the unused branch
of a conditional. In this case, improving the type annotation would lead to a
spurious error: the migrated program would fail even though the original ran without error. 
Although this example is contrived, programs in untyped 
languages often have code whose reachability is environment-dependent
(e.g., JavaScript web programs that support multiple browsers, Python programs
that can be run in Python 2 and 3).
 The flexibility of gradual typing is particularly valuable in
these cases, but reasoning about safety and precision in tandem is subtle. 

\begin{wrapfigure}{r}{0.26\textwidth}
\begin{lstlisting}[aboveskip=-10pt]
  #\tikzmark{C-mark-1}\fbox{int -> int}#      #\tikzmark{C-mark-2}\fbox{int}#
            
let C (f : #\tikzmark{C-mark-3}#STAR) (x : #\tikzmark{C-mark-4}#STAR) =
  if x > 0 then
    1 + f x
  else
    42
\end{lstlisting}
\caption{Context restriction.}
\label{ex-ts-context-restriction}
\begin{tikzpicture}[remember picture,overlay]
  \draw[gray,-latex] ([yshift=5,xshift=3]pic cs:C-mark-3) to [bend right=15] ([yshift=-5,xshift=25]pic cs:C-mark-1);
  \draw[gray,-latex] ([yshift=5,xshift=3]pic cs:C-mark-4) to [bend left=15] ([yshift=-5,xshift=10]pic cs:C-mark-2);
\end{tikzpicture}
\end{wrapfigure}
  
\paragraph{Type migration can restrict the context of a program}

There are many cases where it is impractical to migrate
an entire program at once. For example, the programmer
may not be able to modify the source code of a library;
they may be migrating a library that is used by others;
or it may just be unacceptable to change every file in a large
software project. In these situations, the type migration question is even trickier.

\begin{wrapfigure}{r}{0.34\textwidth}
\begin{lstlisting}[aboveskip=-10pt]
#\tikzmark{D-mark-1}\fbox{int -> int}#

let D (f : STAR#\tikzmark{D-mark-3}#) = 
  f 100 + 10;
  f            #\tikzmark{D-mark-2}\fbox{int -> int}#

let id : STAR#\tikzmark{D-mark-4}# = D(fun (x: STAR) . x)

\end{lstlisting}
\caption{Context restriction.}
\label{ex-ts-context-restriction-2}
\begin{tikzpicture}[remember picture,overlay]
  \draw[gray,-latex] ([yshift=5,xshift=-2]pic cs:D-mark-3) to [bend right=15] ([yshift=-5,xshift=25]pic cs:D-mark-1);
  \draw[gray,-latex] ([yshift=5,xshift=-2]pic cs:D-mark-4) to [bend left=15] ([yshift=5]pic cs:D-mark-2);
\end{tikzpicture}
\end{wrapfigure}

\Cref{ex-ts-context-restriction} shows a higher-order function \lstinline|C|
that calculates $1+f(x)$ when $x$ is greater than zero. We could migrate
\lstinline|C| to require $f$ to be an integer function, which precisely 
captures how \lstinline|C| uses $f$. However, this migration makes 
some calls to \lstinline|C| ill-typed. For example,
\lstinline|C 0 0| evaluates to \lstinline|42| before migration, but is
ill-typed after migration.

\Cref{ex-ts-context-restriction-2} illustrates another subtle interaction
between type-migrated code and its context. The function \lstinline|D|
receives $f$ and expects it to be a function over numbers. Unlike 
the previous example, \lstinline|D| always calls $f$, so it may appear safe
to annotate  $f$ with the type \lstinline|int -> int|.
However, \lstinline|D| also returns $f$ back to its caller, so this migration
changes the return type of \lstinline|D| from \tdyn{} to \lstinline|int -> int|.
 For example, when $f$ is the 
identity function, $D(f)$ returns the identity function before migration,
but after migration $D(f)$ is restricted to only work on \lstinline|int|s. 

\begin{wrapfigure}{r}{0.3\textwidth}
\begin{lstlisting}
let E(id : STAR) =
  id 2;
  id true          #\tikzmark{derr-mig}\fbox{int}#

E(fun (x : #\tikzmark{derr-mark}#STAR) . x);
\end{lstlisting}
\caption{Dynamic type error.}
\label{safe-ts-example}
\begin{tikzpicture}[remember picture,overlay]
  \draw[gray,-latex] ([yshift=5,xshift=3]pic cs:derr-mark) to [bend left=30] ([yshift=2]pic cs:derr-mig);
\end{tikzpicture}
\vspace{-20pt}
\end{wrapfigure}

To summarize, there is a fundamental trade-off between making types
precise in migrated code, and maintaining compatibility with unmigrated code.

\paragraph{Type migration can introduce new dynamic errors}

So far, we have looked at migrations that introduce static type errors. 
However, there is a more insidious problem that can occur: a migration 
can introduce new dynamic type errors. \Cref{safe-ts-example} shows
a program that runs without error: \lstinline|E| receives the identity function
and applies it to two different types. However, since \lstinline|E|'s argument
has type \tdyn, which is compatible with all types, the program is well-typed
even if we migrate the identity function to require an integer argument.
Gradual typing will wrap the function to dynamically check that it only receives
integers.
So the program runs without error before migration, but produces a dynamic type
error after migration. Strictly speaking, although this migration introduces
a new dynamic error, its static types are more precise. 
When evaluating migrations, it is not enough to consider just the types or
interfaces: it is important to understand which run-time checks will be inserted.

\medskip

In summary, there are several competing concerns that we must consider
when choosing an approach to type migration. \ourtool{}
prioritizes preserving the behavior of the original program: it produces types 
that do not introduce new static or 
dynamic errors in the migrated code. However, this objective leaves the
question of context unanswered. Should \ourtool{} produce the most precise
type it can? This may make the migrated code incompatible with unmigrated code.
So, should \ourtool{} instead produce a type that is compatible with all untyped code?
This would mean discarding a lot of useful information, e.g., the types of function arguments.
Or, should \ourtool{} strike a compromise between precision and 
compatibility? We think the right answer depends on the context in which
the type migration tool is being used. Instead of making an arbitrary
decision, \ourtool{} allows the programmer to choose between several
migrations that prioritize different properties.

\section{Formalizing the Type Migration Problem}\label{sec:type_migration}
\label{sec:formalism}

We now formally define the type migration problem. We first briefly review the
\emph{gradually typed lambda calculus} (GTLC)~\cite{siek:refined}, which is a
core calculus for mixing typed and untyped code. We then present
several definitions of type migration for the GTLC.

\begin{figure}
\figsize
\begin{subfigure}{0.33\textwidth}
\figsize
\(
\begin{array}{@{}r@{\,}c@{\,}l@{\quad}l}
\multicolumn{3}{@{}l}{\textbf{Base types}} \\
B & \pdef & \tint \mid \kw{bool} \\
\multicolumn{3}{@{}l}{\textbf{Types and contexts}} \\
S,T & \pdef & B & \textrm{Base type} \\
    & \mid  & S \rightarrow T & \textrm{Function type} \\
    & \mid  & \tdyn & \textrm{Unknown type} \\
\Gamma &\pdef & \cdot \mid \Gamma,x:T \\
\multicolumn{3}{@{}l}{\textbf{Constants}} \\
b & \pdef & \kw{true} \mid \kw{false} & \textrm{Boolean literal} \\
n & \pdef & \cdots & \textrm{Integer literal} \\
c & \pdef & b \mid n \\
\multicolumn{3}{@{}l}{\textbf{Expressions}} \\
e & \pdef & x & \textrm{Identifier} \\
  & \mid & c & \textrm{Literal} \\
  & \mid & \efun{x}{T}{e} & \textrm{Function} \\
  & \mid & e_1~e_2           & \textrm{Application} \\
  & \mid & e_1\times e_2     & \textrm{Multiplication} \\
\end{array}
\)
\end{subfigure}
\vrule
\begin{subfigure}{0.65\textwidth}
\figsize
\(
\begin{array}{l}
\textbf{Type Consistency} \quad \fbox{$T \sim T$} \\
\inferrule*{\phantom{.}}{\tdyn \sim T} \quad
\inferrule*{\phantom{.}}{T \sim \tdyn} \quad
\inferrule*{\phantom{.}}{T \sim T} \quad
\inferrule*{\phantom{.}}{B \sim B} \quad
\inferrule*{S_1 \sim S_2 \\ T_1 \sim T_2}
{S_1 \rightarrow T_1 \sim S_2 \rightarrow T_2} \\[1em]
\textbf{Typing Literals} \quad \fbox{$\mathit{ty}: c \rightarrow B$} \\
\mathit{ty}(n) = \tint \quad \mathit{ty}(b) = \kw{bool} \\[1em]
\textbf{Typing} \quad \fbox{$\Gamma \vdash e : T$} \\[0.5em]
\inferrule*{\Gamma(x) = T}{\Gamma \vdash x : T} 
\quad
\inferrule*{\phantom{.}}{\Gamma \vdash c : \mathit{ty}(c)}
\quad
\inferrule*{\Gamma,x:S \vdash e : T}
{\Gamma \vdash \efun{x}{S}{e} : S \rightarrow T} \\[0.5em]
\inferrule*{\Gamma \vdash e_1 : \tdyn \quad
\Gamma \vdash e_2 : T}
{\Gamma\vdash e_1~e_2 : \tdyn}
\quad
\inferrule*{\Gamma \vdash e_1 : S \rightarrow T \quad
\Gamma \vdash e_2 : S' \quad
S \sim S'}
{\Gamma\vdash e_1~e_2 : T} \\[0.5em]
\inferrule*{\Gamma \vdash e_1 : S \quad
\Gamma \vdash e_2 : T \quad
S \sim \tint \quad T \sim\tint}
{\Gamma\vdash e_1 \times e_2 : \tint} \quad
\end{array}
\)
\end{subfigure}

\caption{The Gradually Typed Lambda Calculus (GTLC): surface syntax and typing.}
\label{fig:gtlc-syntax}
\Description{}
\end{figure}

\subsection{The Gradually Typed Lambda Calculus}\label{subsec:gtlc}

The Gradually Typed Lambda Calculus (GTLC) extends the typed lambda calculus
with base types (integers and booleans) and the \emph{unknown type}
\tdyn. \Cref{fig:gtlc-syntax} shows its syntax and typing rules.

Type checking relies on the \emph{type consistency} relation, $S \sim T$.
Type consistency determines whether an $S$-typed
expression may appear in a $T$-typed context. Two types are consistent if they
are structurally equal up to any unknown (\tdyn) types within them;
the \tdyn-type is consistent with all types and any expression may
appear in a $\tdyn$-typed context. The type
consistency relation is reflexive and symmetric, but not transitive: 
 $\tint$ and $\kw{bool}$ are both consistent with $\tdyn$ but not
  with each other. 

The typing rules for identifiers, literals, and functions are straightforward,
but there are two function application rules: 
(1)~If the expression in function position has type \tdyn,
then the argument may have any type, and the result of the application has type \tdyn.
 (2)~When the type of the function expression is an arrow 
type ($S \rightarrow T$), the result has type $T$. The type of the argument must
be consistent with---but not necessarily equal to---the type of argument the
function expects ($S'\sim S$).

We add a built-in multiplication operator that requires the types of its operands to be 
consistent with \tint (i.e., an operand may have type \tdyn). We choose multiplication 
because the ``+'' operator is overloaded in many untyped languages: we add addition
in \cref{sec:smt_extensions}, where we discuss overloading.

\begin{figure}
\begin{subfigure}{0.31\textwidth}
\figsize
\(
\begin{array}{@{}r@{\,}c@{\,}l@{\quad}l}
\multicolumn{4}{@{}l}{\textbf{Ground types}} \\
G & \pdef & B \mid \kw{fun} \\
\multicolumn{4}{@{}l}{\textbf{Coercions}} \\
k & \pdef & G? & \textrm{Untag} \\
  & \mid & G! & \textrm{Tag} \\
  & \mid & \kw{wrap}(k_1,k_2) & \textrm{Wrap function} \\
  & \mid & k_1;k_2 & \textrm{Sequence} \\
  & \mid & \kw{id}_T & \textrm{Identity} \\
\multicolumn{4}{@{}l}{\textbf{Expressions}} \\
e & \pdef & \cdots \mid [k]\;e & \textrm{Apply coercion} \\
\multicolumn{4}{@{}l}{\textbf{Untagged values}} \\
u & \pdef & \multicolumn{2}{@{}l}{c \mid \efun{x}{T}{e}} \\
\multicolumn{4}{@{}l}{\textbf{Values}} \\
v & \pdef & \multicolumn{2}{@{}l}{u \mid \kw{box}(G,u)} \\ 
\multicolumn{4}{@{}l}{\textbf{Evaluation Contexts}} \\
E & \pdef & \multicolumn{2}{@{}l}{[] \mid E~e \mid v~E \mid [k]\;E}   \\
\multicolumn{4}{@{}l}{\textbf{Active Expressions}} \\
\mathit{ae} & \pdef & \multicolumn{2}{@{}l}{(\efun{x}{T}{e})~v \mid[k]\;v} \\
\end{array}
\)
\(
\begin{array}{@{}l}
\textbf{Evaluation}~\fbox{$\vdash e \hookrightarrow e$} \\[0.5em]
~(\efun{x}{T}{e})~v \hookrightarrow e[x/v] \\
~[\kw{id}]\ v \hookrightarrow v \\
~[G!]\ (u) \hookrightarrow \kw{box}(G,u) \\
~[G?]\ (\kw{box}(G,u)) \hookrightarrow u \\
~[\kw{wrap}(k_1,k_2)]~v \hookrightarrow \\
\qquad \efun{x}{\tdyn}{[k_2]\ (v\ ([k_1]\ x))} \\
~[k_1;k_2]~v \hookrightarrow [k_2]~([k_1]~v) \\
\inferrule*{\mathit{ae} \hookrightarrow e'}{E[\mathit{ae}] \hookrightarrow E[e']}
\end{array}
\)

\end{subfigure}
\vrule
\begin{subfigure}{0.57\textwidth}
\figsize
\(
\begin{array}{l}
\textrm{coerce}(T, T) = \kw{id}_T \\
\textrm{coerce}(\tdyn, B) = B?  \\
\textrm{coerce}(B, \tdyn) = B! \\
\textrm{coerce}(\tdyn,\tdyn\rightarrow\tdyn) = \kw{fun}? \\
\textrm{coerce}(\tdyn\rightarrow\tdyn,\tdyn) = \kw{fun}! \\
\textrm{coerce}(S_1\rightarrow S_2, T_1\rightarrow T_2) = \kw{wrap}(\textrm{coerce}(T_1,S_1),\textrm{coerce}(S_2,T_2)) \\
\textrm{coerce}(\tdyn,T_1\rightarrow T_2) = \kw{fun}?; \kw{wrap}(\textrm{coerce}(T_1,\tdyn),\textrm{coerce}(\tdyn,T_2)) \\
\textrm{coerce}(T_1\rightarrow T_2,\tdyn) = \kw{wrap}(\textrm{coerce}(\tdyn,T_1),\textrm{coerce}(T_2,\tdyn));\kw{fun}! \\
\textrm{coerce}(S,T) = \textrm{coerce}(S,\tdyn);\textrm{coerce}(\tdyn,T) \\[0.5em]
\textbf{Coercion Insertion} \quad \fbox{$\Gamma \vdash e \Rightarrow e, T$} \\[0.5em]
\inferrule*{\Gamma(x) = T}{\Gamma \vdash x \Rightarrow x, T} \quad \inferrule*{\phantom{.}}{\Gamma \vdash c \Rightarrow c, \mathit{ty}(c)} \\[1em]
\inferrule*{\Gamma,x:S\vdash e \Rightarrow e',T}{\Gamma\vdash\efun{x}{S}{e} \Rightarrow \efun{x}{S}{e'},S\rightarrow T} \\[1em]
\inferrule*{\Gamma \vdash e_1 \Rightarrow e_1', S\rightarrow T \quad
\Gamma \vdash e_2 \Rightarrow e_2', S'}  
{\Gamma\vdash e_1\ e_2 \Rightarrow e_1'\ ([\textrm{coerce}(S',S)]\ e_2'), T} \\[1em]
\inferrule*{\Gamma \vdash e_1 \Rightarrow e_1', T \quad T \ne T_{1} \rightarrow T_{2} \quad
\Gamma \vdash e_2 \Rightarrow e_2', S}  
{\Gamma\vdash e_1\ e_2 \Rightarrow ([\textrm{coerce}(T,\tdyn\rightarrow\tdyn)]\ e_1')\ ([\textrm{coerce}(S,\tdyn)]\ e_2'), \tdyn}
\end{array}
\)

\end{subfigure}
\caption{Coercion insertion and evaluation for the GTLC.}
\label{baby-coercion-insertion}
\Description[]

\end{figure}

\subsection{Ground Types and Coercion-based Semantics}\label{subsec:coercion_gtlc}

Programs in the GTLC are not run directly, but are first compiled to
an intermediate representation where static type consistency checks
are turned into dynamic checks if necessary. There are two well-known
mechanisms for describing these dynamic checks: casts and
coercions. We use coercions, following \citet{henglein:coercions}, as they most closely match the type-tagging and tag-checking operations used at run-time in dynamic 
languages.\footnote{The two approaches are inter-translatable~\cite{herman:space,greenberg:thesis} with full abstraction~\cite{siek:coercion}.}

 The \emph{ground types} ($G$ in \cref{baby-coercion-insertion}) are the
types that are dynamically observable, and include all base types and
a ground type \kw{fun} for all functions. The two basic coercions ($k$) 
tag a value with a ground type ($G!$) and untag
a value after checking that it has a particular ground type ($G?$). Both of these
operations can fail: an already-tagged value cannot be re-tagged, and
untagging succeeds only if the value has the expected ground type.
There are three additional coercions: identity coercions, which exist only 
to simplify certain definitions; a sequencing coercion ($k_1;k_2$); and 
a \emph{function proxy} $\kw{wrap}$ that lifts coercions to functions.

To see how the coercion system works, consider a case where
we have a \tdyn-typed value $f$ that we want to treat as a function of type $\tint
\rightarrow \tint$. To do so, we apply $f$ to a coercion as follows:
\[\lbrack \kw{fun}?; \kw{wrap}(\tint!, \tint?)\rbrack f\]
The sequence evaluates from left to right: it first checks that
$f$ is a function ($\kw{fun?}$), 
and then wraps $f$ in a function proxy that will tag its $\tint$ argument
(since $f$ expects a \tdyn value) and will untag its result (since $f$
returns a \tdyn, but we expect an $\tint$).

The values of the language ($v$) include constants, functions, and values
tagged with a ground type. We define tagged values ($\kw{box}(G,u)$) so that a
tag can only be placed on an untagged value ($u$).

The coercion insertion rules are analogous to typing, but produce both a
type and an equivalent expression with explicit coercions.
They rely on the \emetacoerce{} metafunction that translates a static
consistency check $S\sim T$ into a corresponding coercion that is dynamically
checkable. When two types are identical, \emetacoerce{} produces the
identity coercion, which can be safely removed.
The final case of \emetacoerce{} addresses inconsistencies 
($S\not\sim T$). Instead of rejecting programs with inconsistent checks,
we produce a coercion that is \emph{doomed to fail}.
Gradual typing systems sometimes reject programs that demand casts between
incompatible types. However, doing so violates the desired property that migrations
should preserve the behavior of the original program when possible. If we 
rejected these programs, a user would need to excise all incompatibilities, whether
or not they are in live code branches, at the onset of migration.

\begin{figure}
\begin{subfigure}{0.37\textwidth}
\figsize
\(
\begin{array}{@{}l}
\textbf{Type Precision} \quad \fbox{$T \sqsubseteq T$}\\[1em]
\inferrule*{\phantom{.}}{\tdyn \sqsubseteq T} \quad
\inferrule*{\phantom{.}}{T \sqsubseteq T} \quad
\inferrule*{S_1 \sqsubseteq S_2 \\ T_1 \sqsubseteq T_2}
{S_1 \rightarrow T_1 \sqsubseteq S_2 \rightarrow T_2}
\end{array}
\)
\end{subfigure}
\begin{subfigure}{0.62\textwidth}
\figsize
\(
\begin{array}{@{}l}
\textbf{Expression Precision}\quad \fbox{$e \sqsubseteq e$} \\[1em]
\inferrule*{\phantom{.}}{x \sqsubseteq x} \quad
\inferrule*{\phantom{.}}{c \sqsubseteq c} \quad
\inferrule*{e_1 \sqsubseteq e_1' \\ e_2 \sqsubseteq e_2'}
{e_1\ e_2 \sqsubseteq e_1'\ e_2'} \quad
\inferrule*{T \sqsubseteq T' \\ e \sqsubseteq e'}
{\efun{x}{T}{e} \sqsubseteq \efun{x}{T'}{e'}}
\end{array}
\)
\end{subfigure}
\caption{Type and expression precision.}
\label{type-precision}
\Description{}
\end{figure}
  
\subsection{Type Migration}\label{subsec:migration_def}

All formulations of the type migration problem rely on defining \emph{type precision},
where $\tdyn$ is the least precise type. The type precision relation
(\cref{type-precision}), written $S \sqsubseteq T$, is a partial order that
holds when $S$ is less precise than $T$ (or $S$ and $T$ are identical). We use
type precision to define expression precision in the obvious way: an expression is 
more precise than its structural equivalent if its type annotations are more precise
 according to the type precision relation.

\Citet{migeed:decidable} define a type migration as an expression that has more
precise type annotations, and use this definition to study the decidability and
computational complexity of several problems, such as finding migrations
that cannot be made more precise.
\begin{definition}[Type Migration]
\label{def_type_migration}
Given $\vdash e : T$ and $\vdash e' : T'$, $e'$ is a \emph{type migration} of $e$ if
$e \sqsubseteq e'$ and $T \sqsubseteq T'$.
\end{definition}

However, as we argued in \cref{sec:what_matters}, improving type precision
is one of several competing goals for type migration. Another important goal
is to avoid introducing new errors into the program. To reason about this,
we must reformulate the definition of a type migration to relate the 
values produced by the original expression and its migration. 
We propose the following definition of a safe type migration: 

\begin{definition}[Safe Type Migration]
Given $\vdash e : T$ and $\vdash e' : T'$, $e'$ is a \emph{safe
type migration} of $e$ if:
\begin{enumerate}
  \item $e \sqsubseteq e'$;
  \item $T \sqsubseteq T'$; and
  \item $e \hookrightarrow^{*} v$ if and only if $e' \hookrightarrow^{*} v'$ with $v \sqsubseteq v'$.
\end{enumerate}
\end{definition}

This definition of type migration relates the values of the two expressions. 
However, it is too weak. For one thing, it does not say anything about programs 
that produce errors or do not terminate.
But there is a more serious problem: it is too permissive for function types.
For example, given the identity function with type $\tdyn\rightarrow\tdyn$,
this definition allows a type migration that changes its type
to $\tint\rightarrow\tint$, which will produce a dynamic type error
if the function is applied to non-integers.

To address this issue, the definition of type migration must 
take into account the contexts in which the migrated expression may be used.
We define a well-typed program context $C$ as a context with a hole
that can be filled with a well-typed open expression to get a well-typed
closed expression.

\begin{definition}[Well-Typed Program Context]
A program context $C$ is well typed, written  $C : (\Gamma\vdash S)\Rightarrow T$ if
for all expressions $e$ where $\Gamma\vdash e : S$ we have $\vdash C[e]:T$.
\end{definition}

We now define a \emph{context-restricted type migration} as a more
precisely-typed expression that is equivalent to the original
expression in all contexts that can be filled with an expression of
a given type $S$. Note that the type expected by the context ($S$) must be 
consistent (but not identical) with the types of both the original and the migrated expression.

\begin{definition}[Context-restricted Type Migration]
  \label{def:ctx-migration}
Given $\vdash e : T$, $\vdash e' : T'$, and a type $S$ where $S\sim T$ and $S\sim T'$, 
$e'$ is a \emph{context-restricted type migration} of $e$ at type $S$ if:
\begin{enumerate}
  \item $e \sqsubseteq e'$;
  \item $T \sqsubseteq T'$; and
  \item For all $C:(\cdot\vdash S)\Rightarrow U$,  either a)~$C[e]\hookrightarrow^{*} v$ and $C[e']\hookrightarrow^{*} v'$
  with $v \sqsubseteq v'$; b)~both $C[e]$ and $C[e']$ get stuck at a failed 
  coercion;\footnote{
    This definition collapses all errors to stuck states.
    If the GTLC were extended with exception handling, then this definition would
    have to be adjusted.}
   or 
  c)~both $C[e]$ and $C[e']$ do not terminate.
\end{enumerate}
\end{definition}

We call a context-restricted type migration at type \tdyn
a \emph{compatible type migration}.

At the limit, the context's expected type $S$ could be \tdyn, in which case the definition
is essentially equivalent to that of \citet[Theorem 3.22]{rastogi:gti}. However,
this is a very strong requirement that
rules out many informative migrations (\cref{sec:what_matters}). If the programmer
is comfortable making assumptions about how the rest of the program will
interact with the migrated expression, they may choose a more precise $S$, and
 allow a wider range of valid type migrations.

 We present these definitions to describe the type migration problem that we
 seek to address in \ourtool. However, we do not prove that \ourtool{} produces
 a context-restricted type migration. Instead, this paper presents empirical
 results to show the effectiveness of \ourtool{} on the GTLC, when compared to
 other type migration tools.

\section{The \ourtool{} Approach to Type Migration}\label{sec:smt}
\lstset{language=SMTLIB}

We now present \ourtool{}, an approach to type migration that differs in two ways 
from previous work. (1)~Instead of relying on a custom constraint solver,
\ourtool{} produces constraints and an objective function for the Z3 
MaxSMT solver~\cite{z3maxsmt}. (2)~Instead of producing a single migration,
or several migrations without guidance on which to choose,
 \ourtool{} allows the user to choose between migrations that prioritize type precision or
compatibility with untyped code. Moreover,
the \ourtool{} migration algorithm handles these different scenarios
in a uniform way. This section presents \ourtool{}'s type migration
algorithm for the core GTLC. \Cref{sec:smt_extensions} extends
\ourtool{} with additional language features, including some that have not been 
precisely described in prior work.

\begin{figure}
\begin{minipage}{0.39\textwidth}  
\figsize
\(
\begin{array}{r@{\,}c@{\,}l@{\quad}l}
\multicolumn{4}{@{}l}{\textbf{Types}} \\
T                & \pdef & \cdots \\
                 & \mid  & \alpha,\beta,\gamma,\delta 
                         & \textrm{Type metavariables} \\[0.5em]
\multicolumn{4}{@{}l}{\textbf{Coercions}} \\
k                & \pdef & \cdots \\
                 & \mid  & \escoerce(T_1,T_2)
                         & \textrm{Coercion from $T_1$ to $T_2$}
\end{array}      
\)
\vskip 0.5em
\textbf{Type Representation}
\begin{lstlisting}
(declare-datatypes () 
  ((Typ (star) (int) (bool)
        (arr (in Typ) (out Typ)))))
\end{lstlisting}
 
\end{minipage}
\vrule
\begin{minipage}{0.5\textwidth}  
\figsize
\(
\begin{array}{r@{\,}c@{\,}l@{\;}l}
\multicolumn{4}{l}{\textbf{Constraints}} \\
~\phi 
& \pdef & T_1 = T_2            & \textrm{Type equality} \\
& \mid  & w                    & \textrm{Boolean variable (weight)} \\
& \mid  & \phi_1 \wedge \phi_2 & \textrm{Conjunction} \\
& \mid  & \phi_1 \vee \phi_2   & \textrm{Disjunction} \\
& \mid  & \neg\phi             & \textrm{Negation} \\
\end{array}      
\)
\(
\begin{array}{l@{\,}c@{\,}l@{\;}l}
\multicolumn{2}{l}{\textbf{Constraint Metafunctions}} \\
~\eisground \in T \rightarrow \phi \\
~\eisground(T) = T \in B \vee T = \tdyn \rightarrow \tdyn \\[0.5em]
\end{array}      
\)
\end{minipage}
  
\caption{The type constraint language, and language extensions for constraint generation.}
\label{constraint-lang-fig}

\end{figure}

\subsection{The Language of Type Constraints}

For the purpose of constraint generation, we make two additions to
the GTLC (\cref{constraint-lang-fig}):
\begin{enumerate}
  
  \item We extend types with type metavariables ($\alpha$).
  
  \item We introduce a new coercion, $\escoerce(S, T)$, which represents a
  suspended call to the \emetacoerce{} metafunction
  (\cref{baby-coercion-insertion}). The type arguments to \escoerce{} may include
  type metavariables. After constraint solving, we substitute any type metavariables
  with concrete types and use the \emetacoerce{} metafunction to get a primitive coercion ($k$).

\end{enumerate}
Both of these are auxiliary and do not appear in the final program.

The constraints ($\phi$) that we generate are boolean-sorted formulas for a
MaxSMT solver that supports the theory of algebraic datatypes~\cite{smt-adt}.
In addition to the usual propositional connectives, our constraints
involve equalities between types ($T_1 = T_2$) and auxiliary boolean 
variables ($w$). We use these boolean variables to define soft constraints 
that guide the solver towards solutions with fewer non-trivial coercions.

Using Z3's algebraic datatypes, we define a new sort (\texttt{Typ}) that encodes
all types ($T$) except type metavariables. Constraint generation defines a
\texttt{Typ}-sorted constant for every metavariable that occurs in a type. For
example, we can solve the type constraint $\alpha \rightarrow \tint = \beta$
with the following commands to the solver:
\begin{lstlisting}
(declare-const alpha Typ)
(declare-const beta Typ)
(assert (= (arr alpha int) beta))
\end{lstlisting}
This example is satisfiable, and the model assigns \lstinline|alpha|
and \lstinline|beta| to metavariable-free types (represented as \lstinline|Typ|). If 
$\sigma$ is such a model, we write $\textsc{Subst}(\sigma, \beta)$ to mean the
metavariable-free type assigned to $\beta$, i.e., the closure of
substituting with the model $\sigma$.
In this example, $\alpha$ is unconstrained, so there are several possible
models: $\sigma = \{ \alpha \mapsto \kw{int}, \beta \mapsto \alpha
\rightarrow \kw{int} \}$ is one, as is $\sigma' = \{ \alpha \mapsto
\tdyn, \dots \}$. We have $\textsc{Subst}(\sigma, \beta) = \kw{int}
\rightarrow \kw{int}$, while $\textsc{Subst}(\sigma', \beta) = \tdyn
\rightarrow \kw{int}$.

Finally, for succinctness, we define $\eisground(T)$, which produces a constraint
that is satisfiable when $T$ is a ground type.
At the moment, the only ground types are base types and dynamic
function types ($\tdyn\rightarrow\tdyn$). \Cref{sec:smt_extensions} extends the
language with additional types and augments the definition of \eisground.

\begin{figure}

\figsize
\(    
\begin{array}{c}
\multicolumn{1}{@{}l}{\fbox{$\Gamma\vdash e \Rightarrow e,T,\phi$}}  \\

\inferrule*[Left=Id]{
  \phi = (\alpha = \Gamma(x)\wedge w) \vee (\alpha = \tdyn\wedge \neg w) \quad \textrm{$\alpha,w$ is fresh}
}
{\Gamma\vdash x  \Rightarrow [\escoerce(\Gamma(x),\alpha)]x, \alpha, \phi}

\quad\quad\quad\quad

\inferrule*[Left=Const]{
\phi = (\alpha = \mathit{ty}(c)\wedge w) \vee (\alpha = \tdyn\wedge \neg w) \quad \textrm{$\alpha,w$ is fresh}
}
{\Gamma\vdash c \Rightarrow [\escoerce(\mathit{ty}(c),\alpha)]c,\alpha, \phi}\\[1em]

\inferrule*[Left=Fun]
{\Gamma,x:\alpha\vdash e \Rightarrow e', T, \phi_1 \quad
\textrm{$\beta,w$ fresh} \\\\
\phi_2 = (\beta = \alpha\rightarrow T \wedge w) \vee
  (\beta = \tdyn \wedge \eisground(\alpha\rightarrow T) \wedge \neg w)
}
{\Gamma\vdash \kw{fun}(x:\alpha) . e \Rightarrow [\escoerce(\alpha\rightarrow T,\beta)]\kw{fun}(x:\alpha) . e', \beta, \phi_1\wedge\phi_2} \\[1em]

\inferrule*[Left=App]
{\Gamma \vdash e_1 \Rightarrow e_1', T_1, \phi_1 \\
  \Gamma \vdash e_2 \Rightarrow e_2', T_2, \phi_2 \\
  \textrm{$\alpha$, $\beta$, $\gamma$, $w_1$, and $w_2$ are fresh} \\\\
  \phi_3 = (T_1 = \alpha \rightarrow \beta \wedge  w_1) \vee 
          (T_1 = \alpha = \beta = \tdyn \wedge \neg w_1) \\
  \phi_4 = (T_2 = \alpha) \\
  \phi_5 = (\beta = \gamma \wedge w_2) \vee (\gamma = \tdyn \wedge \neg w_2)
}
{\Gamma\vdash e_1~e_2 \Rightarrow 
[\escoerce(\beta,\gamma)](([\escoerce(T_1,\alpha\rightarrow\beta)]e_1')~e_2'), \gamma,
  \phi_1 \wedge \phi_2 \wedge \phi_3 \wedge \phi_4 \wedge \phi_5} \\[1em]

  \inferrule*[Left=Mul]
{\Gamma \vdash e_1 \Rightarrow e_1', T_1, \phi_1 \\
  \Gamma \vdash e_2 \Rightarrow e_2', T_2, \phi_2 \\
  \textrm{$w_1$, $w_2$, and $w_3$ are fresh} \\\\
  \phi_3 = (T_1 = \tint \wedge w_1) \vee 
            (T_1 = \tdyn \wedge \neg w_1) \quad
  \phi_4 = (T_2 = \tint\wedge w_2) \vee 
            (T_2 = \tdyn \wedge \neg w_2) \\\\\
  \phi_5 = (\alpha = \tint \wedge w_3) \vee (\alpha = \tdyn \wedge \neg w_3)
}
{\Gamma\vdash e_1 \times e_2 \Rightarrow 
[\escoerce(\tint,\alpha)]([\escoerce(T_1,\tint)]e_1' \times [\escoerce(T_2,\tint)]e_2'), \alpha,
  \phi_1 \wedge \phi_2 \wedge \phi_3 \wedge \phi_4 \wedge \phi_5}   
  
\end{array}
\)

\caption{Constraint generation for GTLC}
\label{baby-cgen}
\Description{}
\end{figure}

\subsection{Generating Type Constraints}\label{subsec:smt_cg}

We now present constraint generation for the GTLC.
To simplify the presentation, we assume that all bound variables have type
\tdyn{}. Constraint generation is a two-step process:
\begin{enumerate}

  \item We replace every \tdyn{} annotation in the input program with a
  fresh metavariable. The solution to the constraints
  maps these metavariables to types, which may be more precise than \tdyn{}.

  \item We generate constraints by applying deterministic, syntax-directed inference rules.

\end{enumerate}

Since the first step is straightforward, we focus on constraint generation. 
The constraint generation rules are of the form $\Gamma\vdash e \Rightarrow
e',T,\phi$: the inputs are the type environment ($\Gamma$) and
the expression ($e$), and the outputs are as follows:

\begin{enumerate}

  \item An output expression ($e'$) that is equivalent to the input expression,
  but with explicit coercions.
  
  \item A type ($T$), which is the type of the expression, and may include
  metavariables.
  
  \item A constraint ($\phi$) with type-sorted and boolean-sorted 
  free variables.

\end{enumerate}
When formulating constraint generation, there are several requirements to 
keep in mind. First, the constraint $\phi$ may be satisfiable in several ways. 
We will eventually use soft constraints to choose among solutions, but we 
design the constraint generation process so that \emph{all models of $\phi$ 
correspond to valid migrations}. Second, as argued in \cref{sec:what_matters},
 we do not want to reject any programs. We therefore set up constraint 
 generation so that we \emph{do not introduce new static errors}.  
 Our final goal is to \emph{favor informative types}. We do this via soft 
 constraints that penalize the number of non-trivial, syntactic
 coercions. Note that this is not the same as minimizing the 
 number of coercions performed during evaluation, which is a harder 
 problem (but see \citet{campora:herder}).

\paragraph{Constraint Generation Rules}

Constraint generation is syntax directed (\Cref{baby-cgen}),
albeit we assume we can generate fresh names. As 
a general principle, we allow all expressions to be coerced to $\tdyn$:
this enables us to migrate all programs, even though it may generate
coercions that are doomed to fail if they are ever run. This property
is critical to ensure that models exist for all programs 
(Theorem~\ref{thm:completeness}).\footnote{We have also implemented
a version of \ourtool{} that uses an alternative constraint generation rule
for identifiers that enforces rigid types together with a modified version of the 
function application rule that can coerce the function argument.  
This leads to a loss of type precision, but produces type annotations that 
are more robust to code-refactoring. Both approaches are sound and safe at the 
generated types (\Cref{subsec:smt_solving}).}

Following this principle, the rule for identifiers (\textsc{Id}) introduces a 
coercion that is either the identity coercion (when $\alpha$ is $T$, the type of the identifier
 in the environment), or a coercion to $\tdyn$ (when $\alpha$ is $\tdyn$). 
 At a later step (\cref{subsec:smt_solving}), we produce a soft constraint favoring
  $w$ over $\neg w$, which guides the solver towards solutions that avoid the 
  non-trivial coercions when possible.  

Similarly, the rule for constants (\textsc{Const}) generates two new variables: $\alpha$
and a fresh weight variable $w$. The rule constrains the type $\alpha$ to either be 
the type of the constant, or the \tdyn type (i.e, to avoid rejecting $\kw{true} \times 1$). 
In the former case, we constrain $w$ to be true, and in the latter, to false. 
 
The rule for functions (\textsc{Fun}) assumes that the argument is annotated with a
unique metavariable ($\alpha$) and recurs into the function body, which produces
some type $T$. The rule gives the function the type $\beta$ (a fresh metavariable), 
and constrains it to be the type of the function ($\alpha\rightarrow T$) or
the $\tdyn$ type. In the latter case, we also constrain the type of the
function to be the ground type ($\tdyn\rightarrow\tdyn$). We use a weight $w$ to 
prefer the former case without rejecting expressions like $1 \times (\efun{x}{\tdyn}{x})$.

The rule for function applications (\textsc{App}) produces a constraint that is a
conjunction of five clauses: $\phi_1$ and $\phi_2$ are the constraints that arise
when recurring into the two sub-expressions of the application; 
$\phi_3$ constrains the type of the function; $\phi_4$ constrains the type 
of the argument; and $\phi_5$ constrains the type of the result. 
Together, $\phi_3$ and $\phi_4$ capture the two ways
 in which applications can be typed in the GTLC: the function may be of type \tdyn, 
in which case it is coerced to the function ground type, $\tdyn \rightarrow \tdyn$ and $w_1$ is false,
 or the function already has a function type, and $w_1$ is true.
  In either case, the argument type is constrained to be the function input type $\alpha$.
  The final constraint allows the result type, $\beta$, to be coerced to $\tdyn$; 
  $w_2$ is true only if this is a non-trivial coercion.

The rule for multiplication (\textsc{Mul}) produces a
five-part conjunction: $\phi_1$ and $\phi_2$ are the constraints produced
by its operands; $\phi_3$ and $\phi_4$ constrain each operand to either
be $\tint$ or \tdyn and use weights to prefer the former; and $\phi_5$
constrains the type of the result to either be $\tint$ or \tdyn, with
a weight that prefers for the former; again, this is necessary to avoid
rejecting programs.

\paragraph{Example 1: Types for the Identity Function}

Consider the following program, which applies the identity function to $42$ and
\kw{true}, and has the least precise type annotations:\footnote{This is a
variation of the example in \cref{safe-ts-example}.}
\[
  (\efun{id}{\tdyn}{(\efun{n}{\tdyn}{\mathit{id}~\kw{true}})(\mathit{id}~42)})~
  (\efun{x}{\tdyn}{x})
\]
First, consider how we might manually migrate the program. 
One approach is to change the type of $x$ to \tint (underlined below), 
and leave the other annotations unchanged:
\[
  (\efun{id}{\tdyn}{(\efun{n}{\tdyn}{\mathit{id}~\kw{true}})~(\mathit{id}~42)})~
  (\efun{x}{\underline{\tint}}{x})
\]
It is important to note that this program is well-typed and has a more precise
type than the original. However, it produces a run-time type error on
$\mathit{id}~true$, whereas the original program does not.
Fortunately, constraint generation rules out this migration: the outermost 
application coerces the argument type  to \tdyn.
However, the argument type ($\tint\rightarrow\tint$) is not a ground
type, which \textsc{App} also requires.

The following type migration, also constructed manually, is the most precise
migration that does not introduce a run-time error (changes to the original
program are underlined):
\[
  (\efun{id}{\underline{\tdyn\rightarrow\tdyn}}{(\efun{n}{\underline{\tint}}{\mathit{id}~\kw{true}})~(\mathit{id}~42)})~
  (\efun{x}{\tdyn}{x})
\]
However, concluding that $n$ has type \tint requires reasoning about the flow of
 values through the identity function. Our constraint generation rules can't find this 
 solution. Instead, the most precise type allowed by our constraints gives 
 $\mathit{id}$ the type $\tdyn\rightarrow\tdyn$ and leaves $n$ and $x$ at type \tdyn:
\[
  (\efun{id}{\underline{\tdyn\rightarrow\tdyn}}{(\efun{n}{\tdyn}{\mathit{id}~\kw{true}})~(\mathit{id}~42)})~
  (\efun{x}{\tdyn}{x})
\]
This example illustrates an important principle that we follow in 
constraint generation: if we generate a new coercion around an expression
$e$ to type $\tdyn$, then we must also \emph{constrain} the type of $e$ to
be a ground type. As we grow the language with more types, 
 the set of ground types will grow. When this happens, we update the definition 
 of the \eisground predicate, but the rest of constraint generation remains unchanged.

\begin{figure}
\figsize
\begin{algorithmic}[1]
\State \textrm{$\triangleright$ The only annotations in $e$ are $\tdyn$}  
\Function{PreciseMigrate}{e} 
  \State $e_1 \gets \Call{IntroduceMetavars}{e}$ \Comment{Replace every $\tdyn$ with a fresh $\alpha$s}
  \State $\cdot\vdash e \Rightarrow e',T_1,\phi$ \Comment{Generate constraints and objectives}
  \For{$\alpha \in \phi$} \Comment{The set of type metavariables in $\phi$}
    \State \texttt{(declare-const $\alpha$ Typ)}
  \EndFor
  \For{$w \in \phi$} \Comment{The set of weight variables in $\phi$}
    \State \texttt{(declare-const $w$ Bool)}
    \State \texttt{(assert-soft $w$ 1)}
  \EndFor
  \State \texttt{(check-sat $\phi$)} 
  \State $\sigma \gets \texttt{(get-model)}$ \Comment{Model mapping type metavariables to types}
  \State $\textbf{return}~\Call{Subst}{\sigma,e'}$ \Comment{Migrated program with explicit coercions}
\EndFunction
\end{algorithmic}

\caption{Precise Type Migration.}
\label{precise-typ-mig}
\Description{}
\end{figure}

The following theorem establishes that all models that satisfy our constraint 
generation rules produce well-typed expressions.

\begin{theorem}[Type Migration Soundness]
  \label{thm:soundness}
If $\Gamma \vdash e \Rightarrow e', T, \phi$ and $\sigma$ is a model
for $\phi$, then $\textsc{Subst}(\sigma, \Gamma) \vdash \textsc{Subst}(\sigma, e') : \textsc{Subst}(\sigma, T)$.
\ifarxiv
\begin{proof}
  By induction on the coercion insertion judgment (see \Cref{thm:pf:models-wf} for more details).
\end{proof}
\fi
\end{theorem}

\subsection{Solving Constraints for Precise Type Migration}
\label{subsec:smt_solving}

Our formulation of constraint generation produces a constraint
($\phi$) that may have multiple models, all of which encode
valid type migrations of varying precision. Our goal in this section is to
find as precise a migration as possible. To do this, we rely on the MaxSMT 
solver's ability to define \emph{soft constraints}. The solver prefers solutions
that obey these constraints, but can violate them when necessary to produce a model. 

Our constraint generation rules adhere to the following recipe: every rule that introduces
 a coercion also introduces a fresh boolean variable ($w$) that is
\kw{true} when the coercion is trivial ($\mathit{coerce}(T,T)$) and \kw{false}
otherwise. The \textsc{Fun} rule introduces one boolean variable, while
the \textsc{App} rule introduces two, since it may introduce two non-trivial coercions. 

We use the algorithm sketched in \cref{precise-typ-mig}.
For each boolean variable, we produce a soft constraint asserting that $w$ 
should hold (the corresponding coercion should be trivial if possible). 
Given these soft constraints, we check that the formula 
$\phi$ is satisfiable and get a model ($\sigma$) that assigns type metavariables
 to types. We then substitute metavariables with concrete types accordingly.

\paragraph{Example 2: A migration that is too precise}

Consider the following program as an input to our algorithm:
\[
F_1 \triangleq \efun{f}{\tdyn}{\efun{g}{\tdyn}{(f~1) \times (g~f)}}
\]
The algorithm produces the following migration, which has the most
precise types possible:
\[
F_2 \triangleq \efun{f}{\tint\rightarrow\tint}{\efun{g}{(\tint\rightarrow\tint)\rightarrow\tint}{(f~1) \times (g~f)}}
\]
But is the most precise type really the best type? The answer depends
on how the original function was used. For example, in the following context
$F_2$ is not substitutable for $F_1$:
\[
\begin{array}{l@{\quad}|@{\quad}l}
\textrm{Before (produces zero)} & \textrm{After (static type error)} \\
\hline
\underline{F_1}~(\efun{x}{\tdyn}{0})~(\efun{k}{\kw{bool}\rightarrow\tdyn}{k~\kw{true}}) &
\underline{F_2}~(\efun{x}{\tdyn}{0})~(\efun{k}{\kw{bool}\rightarrow\tdyn}{k~\kw{true}})
\end{array}
\]
The left-hand side type-checks and evaluates to $0$, while the right-hand
side has a static type error: the \kw{bool} type in the (unmigrated) context is
inconsistent with the migrated type \tint. 

We might reason that it is acceptable to generate this static error. But 
there is a second, more serious problem: in a gradually typed language, 
it is possible to turn static type errors into run-time type errors. 
Consider the following variation where the annotation on $k$ in the unmigrated 
version is less precise:
\[
\begin{array}{l@{\quad}|@{\quad}l}
\textrm{Before (produces zero)} & \textrm{After (dynamic type error)} \\
\hline
\underline{F_1}~(\efun{x}{\tdyn}{0})~(\efun{k}{\tdyn}{k~\kw{true}}) &
\underline{F_2}~(\efun{x}{\tdyn}{0})~(\efun{k}{\tdyn}{k~\kw{true}})
\end{array}
\]
Both programs above are well-typed. However, the static error from 
the previous example is now a dynamic error. As we argued in \cref{sec:what_matters},
 making types more precise in a portion of a program can introduce 
 run-time errors at the (higher-order) boundary between migrated and unmigrated code.

Perhaps we can address this problem by producing a different migration of $F_1$:
\[
F_3 \triangleq \efun{f}{\tdyn\rightarrow\tint}{\efun{g}{(\tdyn\rightarrow\tint)\rightarrow\tint}{(f~1) \times (g~f)}}
\]
This migration is less precise than $F_2$: although $f$ and $g$ must still be 
functions, they are not required to consume integers. It is therefore
 equivalent to $F_1$ in our unmigrated context.
\[
\begin{array}{l@{\quad}|@{\quad}l}
\textrm{Before (produces zero)} & \textrm{After (also produces zero)} \\
\hline
\underline{F_1}~(\efun{x}{\tdyn}{0})~(\efun{k}{\tdyn}{k~\kw{true}}) &
\underline{F_3}~(\efun{x}{\tdyn}{0})~(\efun{k}{\tdyn}{k~\kw{true}})
\end{array}
\]
Unfortunately, there are other contexts that lead to errors in $F_3$ that
do not occur with $F_1$. For instance, the following program produces an 
error with $F_3$ but not $F_1$.
\[
F_3~(\efun{x}{\tdyn}{x})~(\efun{\mathit{id}}{\tdyn}{(\efun{b}{\tdyn}{0})~(\mathit{id}~\kw{true})})
\]
We can address this problem with a migration with even lower precision:
\[
F_4 \triangleq \efun{f}{\tdyn\rightarrow\tdyn}{\efun{g}{(\tdyn\rightarrow\tdyn)\rightarrow\tint}{(f~1) \times (g~f)}}
\]
This expression does not produce the same error as the previous example,
and is compatible with all our examples. However, we have lost a lot of
information about how $F_1$ uses its arguments. To summarize, we have seen
a series of migrations for $F_1$ in decreasing order of precision:
\[
F_1 \sqsubseteq F_4 \sqsubseteq F_3 \sqsubseteq F_2
\]
Our algorithm produces $F_2$, but the other, less precise migrations are
compatible with more contexts. So, which migration is best? The answer
depends on the context of use for the program. If the programmer is generating 
documentation, they may prefer the more precise migration. On the other hand, 
if they are adding types to a library and cannot make assumptions about the 
function's caller, they may desire the migration that is compatible with more contexts. 

\begin{figure}
\begin{subfigure}{0.55\textwidth}
\figsize
\begin{algorithmic}[1]
\State \textrm{$\triangleright$ The only annotations in $e$ are $\tdyn$}  
\Function{Migrate}{\textsc{Weaken}, e} 
  \State $e_1 \gets \Call{IntroduceMetavars}{e}$ \label{fancy-line-begin}\Comment{Replace $\tdyn$s with fresh $\alpha$s}
  \State $\cdot\vdash e_1 \Rightarrow e',T_1,\phi$
  \For{$\alpha \in \phi$} \Comment{The type metavariables in $\phi$}
    \State \texttt{(declare-const $\alpha$ Typ)}
  \EndFor
  \For{$w \in \phi$} \Comment{The weight variables in $\phi$}
    \State \texttt{(declare-const $w$ Bool)}
    \State \texttt{(assert-soft $w$ 1)}
  \EndFor
  \State \texttt{(check-sat $\phi$)} 
  \State $\sigma \gets \texttt{(get-model)}$
  \State $T_2 \gets \Call{Subst}{\sigma,T_1}$ \label{fancy-line-end-old} \Comment{The most precise type}
  \State $\phi' \gets \Call{Weaken}{T_2,T_1}$\label{fancy-line-call-weaken}
  \State \texttt{(check-sat $\phi \wedge \phi'$)}\label{fancy-line-fancy-constraint}
  \State $\sigma' \gets \texttt{(get-model)}$\label{fancy-line-fancy-model}
  \State $\textbf{return}~ \Call{Subst}{\sigma',e'}$
\EndFunction
\end{algorithmic}
\end{subfigure}
\hfill
\vrule
\hfill
\begin{subfigure}{0.43\textwidth}
\figsize
\(
\begin{array}{@{}r@{\,}c@{\,}l}
\multicolumn{3}{@{}l}{\quad\quad\quad\tikzmark{mig-1}\fbox{$T_1=\tdyn$ \cite{rastogi:gti}}} \\
  & & \\
P(T_1 \rightarrow T_2,\phi_T, b) & \triangleq & \tikzmark{mig-2}\fbox{$P(T_1,\texttt{(arr-in $\phi_T$)},\neg b)$}~\wedge \\
 & & P(T_2,\texttt{(arr-out $\phi_T$)},b) \\
P(B, \phi_T, \kw{true}) & \triangleq & \kw{true} \\
P(B, \phi_T, \kw{false}) & \triangleq & \phi_T =\tdyn \\
P(\tdyn, \phi_T, b) & \triangleq & \kw{true} \\
\textsc{Weaken}(T,\phi_T) & \triangleq & P(T, \phi_T, \kw{true})
\end{array}
\)
\begin{tikzpicture}[remember picture,overlay]
\draw[gray,-latex] ([yshift=9,xshift=15]pic cs:mig-2) to [bend right=10] ([yshift=-5,xshift=35]pic cs:mig-1);
\end{tikzpicture}
\end{subfigure}
\caption{The Type Migration Algorithm.}
\label{fancy-migration}
\Description{}
\end{figure}
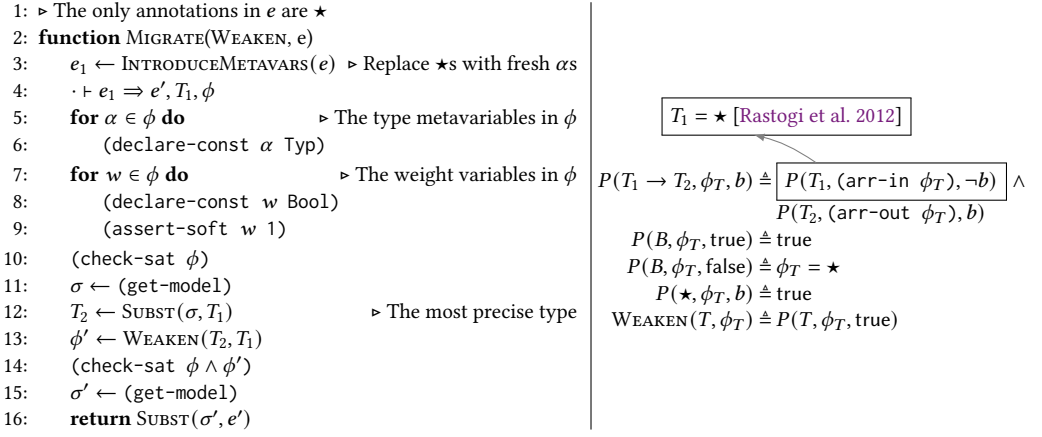

\subsection{Choosing Alternative Migrations}\label{subsec:weaken}

Although the algorithm presented above produces the most precise migration that
the \ourtool constraints encode, we can also use \ourtool to infer alternative 
migrations that prioritize other properties, such as contextual compatibility.

At first glance, it seems straightforward to weaken the more precise type
inferred in the preceding section. Suppose the algorithm produces a
migration $e$ with type $T$, and we want a less precise type $S$ ($S
\sqsubseteq T$). It seems that we could simply wrap $e$ in a coercion:
$[\emetacoerce(T,S)]e$.
Unfortunately, this purported solution is no different from the adversarial 
contexts presented above. The expression has the desired weaker type $S$, 
but gradual typing ensures that it behaves the same as the stronger type $T$ 
at run-time, including producing the same run-time errors! Instead, we need to 
alter the type annotations that are \emph{internal} to $e$.

\paragraph{Weakening Migrations}

\ourtool{} employs a two-step approach to type migration. We first generate
constraints and calculate the most precise type possible ($T_2$), as described earlier
(lines~\ref{fancy-line-begin}--\ref{fancy-line-end-old} of
\cref{fancy-migration}; identical to \cref{precise-typ-mig}).
We then apply the \textsc{Weaken} metafunction, which identifies all
the base types in negative position in $T_2$ (following \citet{rastogi:gti}).
The second argument to \textsc{Weaken} is a \texttt{Typ}-sorted formula
that represents the type of the program ($T_1$).
In \textsc{Weaken}'s helper function $P$, we use this formula to identify
portions of the output type in negative position, and constrain them to
be equal to \tdyn. The result is a constraint ($\phi'$) that weakens the
program type.

Once we have the weakening constraint, we must update the type
annotations in the migrated program and calculate the new weaker type. 
To do so, we run the solver once more with the added constraint
(line~\ref{fancy-line-fancy-constraint}). This produces a new model
(line~\ref{fancy-line-fancy-model}), which we use to substitute type
metavariables and produce a fully annotated program.

It is worth reflecting on why a two-stage procedure is necessary. The first
stage produces the most precise type that we can. This
is necessary to discover a type skeleton that is as precise as possible; otherwise, 
we might miss some of the structure, e.g., by failing to predict arrow types.
The second stage is necessary in order to propagate the constraints on the 
program's type back through the migrated program, which may involve 
arbitrary changes to internal type annotations.

Critically, the new set of constraints $\phi'$ must not impose unnecessary
conditions on the type of the program. For example, suppose the original program $e$
has a precise type $\tint\rightarrow\tint$. Since this type
only allows the context to provide \tint-arguments to $e$, we might
conclude that a better type for $e$ is $\tdyn\rightarrow\tint$. But this
may be impossible: for instance, if $e$ is the identity function, its argument and 
result types must be the same. On the other hand, if the body of $e$ is a multiplication, 
then making the input type \tdyn does not affect the output type: it can remain \tint.
By adding the constraint and re-solving, \ourtool{} is able to distinguish between
these two scenarios.

We note that there are several possible variations for \textsc{Weaken}. 
When migrating higher-order functions, it is useful to use a definition 
that turns base-typed inputs in negative position to \tdyn, but 
preserves arrow types in the input. An alternative is to turn all
input types to \tdyn to maximize compatibility, similar to \citet{rastogi:gti}. 
Our implementation of \ourtool{} supports both of these and could be easily
extended to other variations as well.

Our two-stage approach to contextual safety highlights the key trade-off between
precision and compatibility in type migration. Our first-pass discovers the most
precise types that we can; our second-pass sacrifices some of this
precision to provide compatibility with a wider range of contexts.

\begin{theorem}[Type Migration Completeness]
  \label{thm:completeness}
  Every well scoped dynamic program $e$ has a migration, i.e., there
  exists $e'$, $T$, and $\phi$ such $\cdot \vdash e \Rightarrow e', T,
  \phi$ such that $\phi$ is satisfiable in some model $\sigma$.
  \ifarxiv
  \begin{proof}
    We prove that a fully dynamic model $\sigma$ exists
    (\Cref{thm:dynamic-models-exist}) and that such models are still models after \textsc{Weaken}
    (\Cref{lem:dynamic-models-stable} and
    \Cref{cor:stable-models-exist}).
  \end{proof}
  \fi
\end{theorem}

\begin{figure}
\begin{subfigure}{0.35\textwidth}
\figsize
\(
\begin{array}{@{}r@{\,}c@{\,}l@{\quad\quad}r@{\,}c@{\,}l}
\multicolumn{3}{@{}l}{\textbf{Ground types}} 
& \multicolumn{3}{@{}l}{\textbf{Base Types}}\\
G & \pdef & \cdots \mid \kw{ref} &
B & \pdef & \cdots \mid \kw{unit} \\  
\multicolumn{3}{@{}l}{\textbf{Constants}} 
& \multicolumn{3}{@{}l}{\textbf{Types}} \\
c & \pdef & \cdots \mid \kw{unit} &
T & \pdef & \cdots \mid \tref{T} 
\end{array}
\)

\(
\begin{array}{@{}r@{\,}c@{\,}l@{\quad}l}
\multicolumn{4}{@{}l}{\textbf{Expressions}} \\
e & \pdef & \cdots \\
  & \mid  & \tref{e}      & \textrm{Create cell} \\
  & \mid  & !e               & \textrm{Read cell} \\
  & \mid  & e_1\kw{\pdef}e_2 & \textrm{Write cell} \\
\end{array}
\)
\end{subfigure}
\vrule
\quad
\begin{subfigure}{0.6\textwidth}
\figsize
\(
\begin{array}{l@{\,}c@{\,}l@{\;}l}
\multicolumn{2}{@{}l}{\textbf{Constraint Metafunctions}} \\
\eisground \in T \rightarrow \phi \\
\eisground(T) = T \in B \vee T = \tdyn \rightarrow \tdyn \vee T = \tref{\tdyn} \\[0.5em]
\end{array}     
\) \\
\textbf{Type Representation}
\begin{lstlisting}
(declare-datatypes () 
  ((Typ (star) (int) (bool)
        (ref (to Typ))
        (arr (in Typ) (out Typ)))))
\end{lstlisting}
\end{subfigure}

\vskip 0.5em \hrule \vskip 0.5em

\begin{subfigure}{\textwidth}
\figsize
\(
\begin{array}{c}
\inferrule*[Left=If]
{\Gamma\vdash e_1 \Rightarrow e_1',T_1,\phi_1 \\
  \Gamma\vdash e_2 \Rightarrow e_2',T_2,\phi_2 \\
  \Gamma\vdash e_3 \Rightarrow e_3',T_3,\phi_3 \\
  \textrm{$w_1,w_2,\alpha$ are fresh} \\
  \phi_4 = ((T_1 = \kw{bool}\wedge w_1) \vee (T_1 = \tdyn \wedge \neg w_1)) \wedge
            ((T_2 = T_3 = \alpha \wedge w_2) \vee 
            (\alpha = \tdyn \wedge \eisground(T_2) \wedge \eisground(T_3) \wedge
              \neg w_2))
  }
{\Gamma\vdash \eite{e_1}{e_2}{e_3} \Rightarrow
  \eite{[\escoerce(T_1,\kw{bool})]e_1'}
      {[\escoerce(T_2,\alpha)]e_2'}
      {[\escoerce(T_3,\alpha)]e_3'},
  \alpha,\phi_1\wedge\phi_2\wedge\phi_3\wedge\phi_4}
\\[1em]
  \inferrule*[Left=Add]
{\Gamma\vdash e_1 \Rightarrow T_1,e_1' \phi_1 \\
  \Gamma\vdash e_2 \Rightarrow T_2,e_2', \phi_2 \\
  \textrm{$w,\alpha$ are fresh} \\\\
  \phi_3 = ((\alpha = \tint \vee \alpha = \kw{str}) \wedge 
            \alpha = T_1 = T_2 \wedge
            w) \vee
          (\alpha = \tdyn \wedge \eisground(T_1) \wedge \eisground(T_2) \wedge \neg w)
}
{\Gamma\vdash e_1+e_2 \Rightarrow 
  ([\escoerce(T_1,\alpha)]e_1') + ([\escoerce(T_2,\alpha)]e_2'), 
  \alpha,\phi_1\wedge\phi_2\wedge\phi_3}\\[1em]
  \inferrule*[Left=Ref]
{\Gamma\vdash e \Rightarrow T, e', \phi_1\\
\textrm{$\alpha$ and $w$ are fresh}\\\\
\phi_2 = (\alpha = \tref{T} \wedge w) \vee (\alpha = \tdyn \wedge \eisground(\tref{T}) \wedge \neg w)}
{\Gamma\vdash \tref{e}\Rightarrow [\escoerce(\tref{T},\alpha)]\tref{e'}, \alpha, \phi_1 \wedge \phi_2}\\[1em]\inferrule*[Left=Deref]
{\Gamma\vdash e \Rightarrow T, e', \phi_1 \quad
  \textrm{$\alpha,w$ are fresh} \quad
  \phi_2 = (T=\tref{\alpha} \wedge w) \vee (T = \alpha = \tdyn \wedge \neg w)}
{\Gamma\vdash !e\Rightarrow !([\escoerce(T,\tref{\alpha})]e'), \alpha,\phi_1 \wedge \phi_2}
\\[1em]
\inferrule*[Left=SetRef]
{\Gamma\vdash e_1 \Rightarrow T_1,e_1',\phi_1 \\
  \Gamma\vdash e_2 \Rightarrow T_2,e_2',\phi_2 \\
  \textrm{$\alpha$ and $w$ are fresh} \\\\
  \phi_3 = (T_1=\tref{\alpha} \wedge T_2 = \alpha \wedge w) \vee
          (\alpha = \tdyn \wedge \eisground(T_1) \wedge \eisground(T_2) \wedge \neg w)}
{\Gamma\vdash e_1\kw{\pdef}e_2\Rightarrow 
  ([\escoerce(T_1,\tref{\alpha})]e_1')\kw{\pdef}[\mathit{coerce}(T_2,\alpha)]e_2', \kw{Unit},
  \phi_1\wedge\phi_2\wedge\phi_3}
\end{array} 
\)
\end{subfigure}

\caption{Extensions to the GTLC.}
\label{fig:mig-exts}
\Description{}
\end{figure}

\section{Language Extensions}
\label{sec:smt_extensions}

We now extend the GTLC and \ourtool{} to support several common language
features. These new features affect our constraint generation rules, but they
do not  change the migration algorithm.

\paragraph{Conditionals}
Retrofitted type checkers for untyped languages employ a variety of techniques
to give precise types to conditional expressions (\cref{sec:related}). The
 GTLC-based languages (e.g., \citet{kuhlenschmidt:grift})
use a simpler approach: (1)~the type
of the test must be consistent with \kw{bool}, and (2)~the type of the
expression is the least upper bound of the types of either branch.

The \textsc{If} rule in \cref{fig:mig-exts} shows constraint generation for
conditionals. The generated constraint ($\phi_4$) has two
conjunctions that 1)~constrain the type of the condition to \kw{bool} or
\tdyn, and 2)~constrain the types of each branch to be identical types or
distinct ground types (in which case, both are coerced to the unknown type). 

\paragraph{Overloaded Operators}

Many languages have overloaded built-in operators: for instance, the ``+'' operator 
is frequently used for addition and string concatenation. To support this, the
run-time system has three operators available: (1)~primitive addition, (2)~primitive
string concatenation, and (3)~a complex operation whose behavior depends on 
the run-time types of its arguments. Type migration can reveal
the type at which an overloaded operator is used, which can help programmers
understand their code and improve run-time performance.
The constraint generation rule for ``+'' in \cref{fig:mig-exts} introduces a 
boolean-sorted variable ($w$) that is true when the operands both have type \tint or
\kw{str}; when the variable is false, the constraint requires the two arguments
to have type \tdyn. Thus, it favors solutions that do not employ \tdyn{} when possible.

\paragraph{Mutable Data Structures}

\ourtool{} supports ML-style mutable references and mutable vectors.
There are several ways to add mutable references to the 
GTLC~\cite{siek:gtlc,herman:space,gtlc-mono}. However, all approaches 
share the following property: in untyped code, where all mutable cells contain 
\tdyn-typed values, the \emph{only} reason that reading or writing fails is when
 the expression in reference position is not a reference. In constraint 
generation, we are careful to avoid solutions that may introduce other kinds of errors.

The least precise reference type is a reference to the unknown type
($\tref{\tdyn}$), so we add this to the set of ground types (\cref{fig:mig-exts}). In
the constraint generation rule for writes, we require that either 
(1)~the type of value written is exactly the referenced type, or
(2)~both the reference and the value written are ground types. 
The restriction to ground types is necessary because, as in the function case, 
once a reference is coerced to $\tdyn$, we have no way to recover its original type;
allowing non-ground types to be coerced to $\tdyn$ can introduce run-time errors. 
\ourtool{} also supports mutable vectors implemented along the same lines.

\paragraph{Other language features} The implementation of \ourtool{} supports a
variety of other language features, including tuples, \texttt{let}, and a
\texttt{fix} construct. Many of these are necessary to support the
Grift programming language, which we use in our evaluation. Constraint generation 
rules for these extensions can be found in
\ifarxiv \cref{cg_appendix} \else [FILL] the submitted supplement (will be on ArXiV in the camera-ready). \fi

\section{Evaluation}\label{sec:evaluation}

This section presents the first comprehensive comparison of several type
migration algorithms from the literature (along with \ourtool). We first compare
\numtools{} type migration tools on a suite of \numbenchmarks{}
programs, including several new benchmarks. We also evaluate
\ourtool{} using the Grift benchmarks from \citet{kuhlenschmidt:grift} to show
that \ourtool can reconstruct hand-written type annotations in Grift.

\begin{figure}
\figsize
\(
\begin{array}{@{}l|l}
\textrm{Name} & \textrm{Expression} \\
\hline
\textrm{\textsc{FArg-Mismatch}}
& (\efun{f}{\tdyn}{f~\kw{true}})~(\efun{x}{\tdyn}{x + 1})
\\
\textrm{\textsc{Rank2-Poly-Id}}
& (\efun{i}{\tdyn}{(\efun{a}{\tdyn}{(i~\kw{true})})~(i~5)})~(\efun{x}{\tdyn}{x})
\\
\textrm{\textsc{Unreachable-Err}}
& (\efun{b}{\tdyn}{b~(\efun{c}{\tdyn}{(\efun{x}{tdyn}{x~x})~5~5})~(\efun{d}{\tdyn}{0})})~(\efun{t}{\tdyn}{\efun{f}{\tdyn}{f}})
\\
\textrm{\textsc{F-In-F-Out}*}
& (\efun{f}{\tdyn}{(\efun{y}{\tdyn}{f})~(f~5)})~(\efun{x}{\tdyn}{10+x}) \\
\textrm{\textsc{Order3-Fun}*}
& \phantom{(}\efun{f}{\tdyn}{\efun{x}{\tdyn}{x~(f~x)}} \\
\textrm{\textsc{Order3-IntFun}*}
& \phantom{(}\efun{f}{\tdyn}{\efun{g}{\tdyn}{f~g~((g~10) + 1)}} \\
\textrm{\textsc{Double-F}*}
& \phantom{(}\efun{f}{\tdyn}{f~(f~\kw{true})} \\
\textrm{\textsc{Outflows}*}
& (\efun{x}{\tdyn}{x~5+x})~5 \\
\textrm{\textsc{Precision-Relation}*}
& (\efun{f}{\tdyn}{f~\kw{true}+(\efun{g}{\tdyn}{g~5})~f})~(\efun{x}{\tdyn}{5}) \\
\textrm{\textsc{If-Tag}}
& \phantom{(}\efun{\mathit{tag}}{\tdyn}{\efun{\mathit{x}}{\tdyn}{
  \eite{\mathit{tag}}{x+1}{\eite{x}{1}{0}}}}
\end{array}
\)

\caption{Our Type Migration Challenge.}
\label{challenge-set}
\Description{}
\end{figure}

\subsection{Gradual Type Migration Benchmarks}
\label{subsubsec:challenge_set}

We evaluate type migration tools using a two-part benchmark suite: a suite of benchmarks
from \Citet{migeed:decidable}, and a new suite of \emph{challenge programs}
designed to illustrate the strengths and weaknesses of different
approaches to type migration. 
Our proposed challenge suite is presented in \cref{challenge-set}. We describe
the ten programs below. Although \ourtool{} supports several extensions 
to the GTLC (\cref{sec:smt_extensions}), we do not use them in the challenge
suite so that we can run as many tools as possible. (The final \textsc{If-Tag}
benchmark is an exception.)

\begin{enumerate}

\item \textsc{FArg-Mismatch}: crashes at run-time, because the functional argument $f$
expects an integer, but is applied to a boolean.

\item \textsc{Rank2-Poly-Id} (based on \cref{safe-ts-example}): defines the
identity function and applies it to a number and a boolean. It uses a Church
encoding of \kw{let}-binding and sequencing that would require rank-2
polymorphism in an ML dialect.

\item \textsc{Unreachable-Err} (based on \cref{ex-ts-unreachable}): has a
crashing expression similar to \textsc{FArg-Mismatch}, but it is unreachable.
The example encodes a conditional as a Church boolean.

\item \textsc{F-In-F-Out}: defines a local function $f$ that escapes.

\item \textsc{Order3-Fun}: a higher-order function that receives
two functions $f$ and $x$. Moreover, the body calculates $f~x$, so
$f$ must be a higher-order function itself.

\item \textsc{Order3-IntFun}: similar to \textsc{Order3-Fun}, but
the program uses operations that force several types to be \tint.

\item \textsc{Double-F}: calculates $f~(f~\kw{true})$. The inner application
suggests that $f$'s argument must be \kw{bool}. However, that would rule out
$\efun{x}{\tdyn}{0}$ as a possible value for $f$.

\item \textsc{Outflows}: defines a function that uses its argument
as two different types. However, the function receives an integer.

\item \textsc{Precision-Relation}: names a function $f$ that must receive
$\tdyn$, since $f$ is applied twice to two different types. However,
the second application re-binds $f$ to $g$, thus $g$ may have a more
precise type.

\item \textsc{If-Tag}: receives a boolean and uses its value to determine the
type of $x$. Conditionals are not in the core GTLC and not supported by all the
tools that we consider. However, it is essential to think through conditionals,
since they induce a type constraint between both branches, and a Church
encoding incurs a significant loss of precision.

\end{enumerate}

Some of these programs (marked with an asterisk in \cref{challenge-set}) can be
given types using Hindley-Milner type inference via translation into OCaml
or Haskell. Doing so reveals important differences between conventional static 
types and the GTLC. For example, the most general type of \textsc{Order3-Fun} 
is a type scheme with two type variables. The GTLC does not support 
polymorphism, so a type migration must use \tdyn{} rather than the more precise
  type. In contrast, the type of $f$ in \textsc{Double-F} is $\kw{bool}\rightarrow\kw{bool}$. 
  However, $f$ can have other types in the GTLC.

\subsection{Benchmarked Type Migration Tools}
We evaluate the performance of the following tools, which have
a variety of different goals, which we describe below:

\begin{enumerate}

\item \ourtool{}: our tool, which we run in two modes: (a)~\ourtoolp produces
a safe migration, and (b)~\ourtoolc produces a compatible migration.
In both modes, \ourtool{} maximizes precision, and migrates all closed
programs.

\item \textsc{Gtubi}: \emph{gradual typing with unification-based
inference}~\cite{siek:gti} is the earliest work on gradual type
migration. It produces safe migrations.

\item \textsc{InsAndOuts}: our implementation of the algorithm in
\citet{rastogi:gti}. The algorithm produces compatible migrations.

\item \textsc{MaxMigrate}: \citet{migeed:decidable} presents algorithms for
several migration problems. We use the \emph{maximal migration} tool, which
produces a migration that cannot be made more precise. The tool searches 
for migrations by building types up to some depth (we use depth five as 
in the paper). A single program may have several maximal migrations; we take
the first migration the tool produces. We halt with no output if no migration 
is found.

\item \textsc{MGT}: our implementation of the algorithm in
\citet{campora:migrating} for migrating untyped or partially typed programs. We
start from untyped code (all functions annotated with $\tdyn$), and take the 
first migration it produces.

\end{enumerate}

\begin{figure}
\figsize
\begin{tabular}{|l|r|r|r|r|}
\hline
Tool
& Migrations
& Safe Migrations
& Compatible Migrations
& Improved Type Annotations \\
& (\% of programs) & (\% of programs) & (\% of programs) & (\% of annotations) \\
\hline
\toolgtubi{} & 0.36 & 0.36 & 0.32 & 0.76 \\ 
\toolrastogi{} & 0.91 & 0.91 & 0.91 & 0.43 \\ 
\toolcampora{} & 1.00 & 1.00 & 0.86 & 0.48 \\ 
\toolmigeed{} & 0.77 & 0.64 & 0.32 & 0.73 \\ 
\ourtoolc{} & 1.00 & 1.00 & 1.00 & 0.31 \\ 
\ourtoolp{} & 1.00 & 1.00 & 0.86 & 0.57 \\ 
\hline
\end{tabular}
\caption{Concise performance metrics for automated type migration tools
(higher numbers are better).}
\label{concise-summary}
\Description{}
\end{figure}

\subsection{Concise Evaluation}

Using our suite of benchmarks, \Cref{concise-summary} shows how the
aforementioned tools perform on the axes of safety, compatibility, and
precision. The first column of numbers reports the percentage of programs
that are successfully migrated, and it is important to take this column
into consideration when interpreting the other columns. For example, the
final column suggests that the \toolgtubi{} outperforms \toolmigeed{} on
type precision. However, the first column shows that \toolgtubi{} only migrates
half as many programs as \toolmigeed{}.

\subsection{How Should Type Migration Tools Be Evaluated?}\label{subsec:how_eval}

The concise evaluation masks many subtle issues that arise in type migration.
For instance, using the total number of type
 annotations improved is a good metric for type precision, but reporting only precision
 obscures the fact that not all improvements are alike: some change the 
 behavior of the original program, while others preserve its semantics. We have also 
 illustrated how type precision can come at the expense of compatibility with unmigrated 
 code. This sacrifice may sometimes be warranted, but when a function is 
 migrated, it should remain usable with at least \emph{some arguments}. 
This seems like a trivial point, but consider the following migration:
\[
\begin{array}{l@{\quad}|@{\quad}l}
\textrm{Original Program} & \textrm{Migrated Program} \\
\hline
\efun{f}{\tdyn}{\efun{x}{\tdyn}{f x x}} &
\efun{f}{\tint\rightarrow\kw{bool}\rightarrow\tdyn}{\efun{x}{\tdyn}{f x x}}
\end{array}
\]
The migrated program has types that cannot be made
more precise. However, the type of $f$ requires $x$ to be both an integer
and a boolean, and thus renders the function unusable.

We propose a multi-stage evaluation process for automated type migration tools. 
For each tool, (1)~we start with the full suite of programs and ask, \emph{How many
programs does the tool reject with static errors?} (2)~We take the
\emph{remaining programs} and ask, \emph{How many migrated programs crash with
a new dynamic type error?} (3)~We take the \emph{remaining programs}
and ask, \emph{How many migrated programs are functions that are rendered unusable?}
(4)~We take the remaining programs and ask two final questions: (a)~\emph{How many
migrated programs are functions with types that are incompatible with some untyped
contexts?} and (b)~\emph{How many type
annotations, counted across all remaining programs, are not improved by
migration?}

For our evaluation, we partially automate the multi-step process described
above. To trigger errors, benchmarks that are functions require an input, which
we construct manually. For step~3, we inspect every program that
crashes on some input, to determine if there is any other input that will make
it not crash. In step~4a, to label a benchmark as incompatible with some contexts,
we manually provide a context  that leads
to a dynamic error, and the benchmarking framework verifies that the error
definitely occurs.  Conversely, to label a benchmark as compatible with all
contexts, we provide a hand-written, compatible migration and the benchmarking
framework verifies that the migration returned by the tool is less precise than
the hand-written migration.

Note that the denominator (potentially) decreases at each stage: if a tool fails
to migrate a program, then it is impossible to assess whether the migrated 
program crashes with a dynamic error. Moreover, we do not want to give a system
credit for increasing the precision of a type if the refinement triggers a new 
dynamic error (i.e., it was an unsafe migration).

\begin{figure}
\figsize
\begin{tabular}{l|r|r|r|r||r}
\cline{2-6}
& \multicolumn{5}{l}{The tool rejects the program, e.g., $1 + \kw{true}$} \\[0.5em] \cline{3-6}
& &  \multicolumn{4}{l}{$(\efun{\mathit{id}}{\tdyn}{\mathit{id}~1})~(\efun{x}{\kw{bool}}{x})$ crashes} \\[0.5em] \cline{4-6}
& & & \multicolumn{3}{l}{$(\efun{f}{\tint\rightarrow\kw{bool}\rightarrow\tdyn}{\efun{x}{\tdyn}{f x x}}$ is unusable} \\[0.5em] \cline{5-6}
& & & & \multicolumn{2}{l}{$\efun{x}{\tint}{x}$ is restricted } \\[0.5em] \cline{6-6}
Tool 
& \textFrac{Rejected}{Total Programs} 
& \textFrac{New Dynamic Errors}{Remaining Programs} 
& \textFrac{Unusable Functions}{Remaining Programs} 
& \textFrac{Restricted Functions}{Remaining Programs} 
& \textFrac{Not Improved}{Total}\tdyn \\[0.5em]
\hline
\toolgtubi & 14 / 22 & 0 / 8 & 0 / 8 &  1 / 8 & 4 / 17 \\
\toolrastogi & 2 / 22 & 0 / 20 & 0 / 20 &  0 / 20 & 24 / 42 \\
\toolcampora & 0 / 22 & 0 / 22 & 0 / 22 &  3 / 22 & 30 / 58 \\
\toolmigeed & 5 / 22 & 3 / 17 & 3 / 14 &  4 / 11 & 6 / 18 \\ 
\ourtoolc & 0 / 22 & 0 / 22 & 0 / 22 &  0 / 22 & 40 / 58 \\
\ourtoolp & 0 / 22 & 0 / 22 & 0 / 22 &  3 / 22 & 25 / 58 \\
\end{tabular}

\caption{Summary of type migration results. \ourtoolc{} favors compatibility
with unmigrated code, and \ourtoolp{} favors precision. Above each column, we show an
example of the kind of migrated program we count in that column.
\textbf{Warning:}
This figure requires careful interpretation. For example, the
\emph{Restricted Functions} column shows that both \toolrastogi{} and
\ourtoolc{} produce zero restricted migrations. However, the denominator is not
the same: \toolrastogi rejects two programs in the first column and only 20
programs remain. In contrast, \ourtoolc{} runs on the full suite of 22 programs.
Thus, \ourtoolc{} is arguably better than \toolrastogi{} since it does not
reject any programs. Furthermore, consider \toolgtubi{} and \toolmigeed{},
both of which can produce restricted migrations. The penultimate column
shows that \toolgtubi{} only produces one restricted program, whereas
\toolmigeed{} produces three, which suggests that \toolgtubi{} is better.
However, \toolgtubi{} statically rejects far more programs in the first column,
thus more programs remain for \toolmigeed{}.}

\label{eval-summary}
\Description{}
\end{figure}
  
\subsection{Comprehensive Evaluation}

The results of our evaluation illustrate the various strengths and 
weaknesses of different approaches to automated type migration. Before diving
into the details of the complete results, we present a bird's-eye view of the
evaluation scheme proposed in \cref{subsec:how_eval}.

\paragraph{How to Interpret \cref{eval-summary}} The summary table in
\cref{eval-summary} must be interpreted carefully. Every row in the table
shows the results of a single tool. The columns can be interpreted as follows:
\begin{itemize}

  \item The rightmost column counts the number of type annotations that are
  \emph{not improved} (i.e., lower is better), out of the total number of
  annotations in the migrated programs that are either context-restricted,
  or compatible with all contexts. Thus we do not count type annotation
  improvements that lead to immediate crashes or render the function unusable.
  Thus, \emph{if safety and precision are the primary concern, this
  column is most informative}.

  \item The first three columns count programs that fail to migrate safely
  in three ways: programs that are rejected statically, programs that crash
  after migration with new dynamic errors, and programs that become unusable
  functions after migration.

  \item In the penultimate column, the denominator counts programs that 
  migrate safely without the errors mentioned above, and the numerator counts 
  programs that still produce errors in some contexts. Therefore, if
  \emph{safety is the only concern}, then the denominator of the penultimate
  column (higher values are better) is the most informative.  On the other
  hand, if \emph{safety and compatibility are both significant}, then the
  difference between the denominator and numerator is most informative
  (a higher value is better).

\end{itemize}

Thus this figure has a deliberate safety bias: it discounts rejected programs,
and increasing type precision, when doing so introduces errors. However, 
the figure makes it possible to compare tools on three axes: 
precision (conditioned on safety), compatibility (which is
meaningless without safety), and safety alone.

\paragraph{Discussion of \cref{eval-summary}}

We include results from running \ourtool{} in two different modes: \ourtoolp{} prioritizes 
precision, while \ourtoolc{} prioritizes contextual compatibility. By design, \ourtool{}
does not produce static or dynamic errors. When it is configured for type precision
 (\ourtoolp), it does restrict the inputs of three functions. However, even in this mode, 
the remaining 19 programs remain compatible with all callers. On the other hand,
when it is configured to prioritize contextual compatibility (\ourtoolc), no
programs are restricted, but fewer types are improved.

Like \ourtool{}, \toolcampora{} restricts some functions, but does not 
produce static or dynamic errors. \toolgtubi{} rejects several programs
statically and restricts the behavior of some functions. However, it does not
introduce any dynamic errors. \toolmigeed{} rejects a few programs: some 
do not have maximal migrations, on others it cannot find a
migration within its search space, and one of our programs uses a conditional,
which is unsupported. In addition, the tool introduces run-time errors in some
programs, and makes some functions unusable. \toolrastogi{} rejects two programs.\footnote{These are two programs from the \citet{migeed:decidable} benchmarks. From correspondence
with the authors of \citet{rastogi:gti}, our implementation seems faithful to the 
presentation in the paper, and the original implementation for Adobe ActionScript is no
longer accessible.} On the remaining programs, it produces migrations that
are compatible with arbitrary unmigrated code as intended. In fact, when we
prioritize compatibility with unmigrated code, \toolrastogi{} outperforms
all other approaches.

The right-most column of the table reports the number of type annotations that
are \emph{not} improved, and this must be interpreted very carefully. The point
of gradual typing is that \tdyn{} serves as an ``escape hatch'' for programs
that cannot be given more precise types. Our suite includes programs that \emph{must}
have some \tdyn{}s, so every tool will have to leave some \tdyn{}s unchanged.
We naturally want a tool to improve as many types as possible, so we may prefer
a tool that has the fewest number of unimproved types. However, notice that
the denominator varies considerably. For example, \ourtoolp{}
cannot improve about half the annotations, but it does not introduce any
errors. In contrast, the oldest tool---\toolgtubi{}--only leaves a small
fraction of annotations unimproved, but it statically rejects the majority of
programs.

\begin{figure}
\figsize
\input{challenge_results}
\caption{Migrations of the challenge set with \ourtool in precise mode.}
\label{challenge-results}
\Description{}
\end{figure}

\paragraph{Challenge Set Results}

We now examine performance on the challenge set in more detail.
\Cref{challenge-results} shows the migrated challenge programs produced by
these tools. We present and discuss results produced by running \ourtool 
to prioritize precision (\ourtoolp); results from \ourtoolc can be found in the appendix.

Examining the detailed output on the challenge set programs reveals interesting
differences in the migrations inferred by the various type migration tools, 
reflecting their differing priorities.

\begin{enumerate}

  \item \textsc{FArg-Mismatch}: \toolrastogi{} produces
  the most precise and informative result, showing that $x$ is a boolean next to
  $x + 100$, which helps locate the error in the program.

  \item \textsc{Rank2-Poly-Id}: \toolrastogi{} and \ourtool{}
  produce the best result that does not introduce a run-time error.
  \toolmigeed{} produces the most precise static type, but has a
  dynamic type error.

  \item \textsc{Unreachable-Err}: \ourtool{}, \toolcampora{}, and \toolrastogi{} are the only
  tools that produce a result. The erroneous and unreachable portion gets
   the type \tdyn{} in \ourtool{}; whereas \toolrastogi{} produces a
  type variable. The rest of the program has informative types.

  \item \textsc{F-In-F-Out}: \toolcampora{}, \toolgtubi, and \ourtool{}
  produce the most precise result. \toolmigeed{} produces an alternative, 
  equally precise type, but introduces a dynamic type error.

  \item \textsc{Order3-Fun}:
  \toolgtubi{} produces the best result. Its result has type variables, thus
  is a type scheme. However, in a larger context,
  these variables would unify with concrete GTLC types.
  \toolcampora{} and \ourtool{} produce a similar result, but with \tdyn.
  \toolmigeed{} produces $\tint\rightarrow\tint$ as the type of $f$, which
  is maximal, but introduces a subtle problem: $(f~x)$ requires $x$
  to be an integer, but $x~(f~x)$ requires $x$ to be a function. 
  
  \item \textsc{Order3-IntFun}: the results are similar to \textsc{Order3-Fun},
  with \toolgtubi{} again doing the best. However, since the program forces certain
  types to be \tint, \ourtool{} and \toolcampora{} now produce the same result.
  
  \item \textsc{Double-F}: \toolmigeed{} produces the best result.
  The most informative annotation on $f$ that is compatible with all 
  contexts is $\tdyn\rightarrow\tdyn$; no tool produces this type. 

  \item \textsc{Outflows}: \toolrastogi{} and \ourtool{} produce the best result.
  This program requires $x$ to have two different types and thus crashes.
  Because the function receives an integer for $x$, \citet{rastogi:gti} and \ourtool{}
  give $x$ the type \tint. The other tools are not capable of reasoning in this manner.
  In a modification of this example where $x$ is used with different 
   types in each branch of a conditional, all tools would likely produce 
   similar results.

  \item \textsc{Precision-Relation}: \toolrastogi{} produces the most
  precise type that does not introduce a run-time type inconsistency.
  \ourtool{} does not give $g$ the most precise type; \toolcampora{} 
  does not improve the type of $f$; and \toolmigeed{} finds a maximal 
  migration that constrains $f$'s argument to \kw{bool}.
  
  \item \textsc{If-Tag}: \toolgtubi{} and \toolmigeed{} do not support 
  conditionals. \ourtoolp{} and \toolcampora{ } produce an unusual result that
  restricts the
  type of the argument $x$ to \kw{bool} and turns the $x+1$
  into $([\kw{bool}!]x)+1$. If we were migrating a larger program that had
  this function as a sub-expression, and this function were actually applied
  to values of $x$ with different types, the type of $x$ would be $\tdyn$.

\end{enumerate}

\paragraph{\Citet{migeed:decidable} Benchmarks} 

\Citet{migeed:decidable} compare their maximal migration tool to the type
migration tool in \citet{campora:migrating}. We extend
the comparison to include \ourtool, \toolgtubi, and \toolrastogi. The
artifact that accompanies this paper includes the complete suite of benchmark
results, and we include all of these
benchmarks in our summary (\cref{eval-summary}).

\paragraph{Summary}

Our type migration challenge suite is designed to highlight
the strengths and weaknesses of different algorithms. As discussed
in \cref{sec:what_matters}, the competing goals of the type migration problem lead
to a range of compromises; we do not claim that any one approach is best, since 
each approach reflects a different weighting of priorities. Because our challenge programs are
synthetic, it would be possible to build a large set of programs that favor one tool
at the expense of others. Our goal has been instead to curate a small set that 
illustrates a variety of weaknesses in every tool.  In addition, our challenge programs are unlikely 
to be representative of real-world type migration problems. A more 
thorough evaluation would require scaling type migration tools to a widely-used 
language with a corpus of third-party code, which is beyond the scope of this paper.

\subsection{Grift Performance Benchmarks}
\label{grift-benchmarks}

\citet{kuhlenschmidt:grift} present a benchmark suite to evaluate the performance 
of Grift programs (running time and space efficiency). Grift extends the GTLC 
with floating-point numbers, characters, loops, recursive functions, tuples, 
mutable references, vectors, and several primitive operators.
Each benchmark has two versions: an untyped version and a fully-typed, hand-annotated
 version. We use \ourtool{} (in precise mode) to migrate every untyped benchmark, 
 and compare the result to the human type annotations. \ourtool supports all 
Grift features except equirecursive types. However, because Grift's 
 equirecursive types do not introduce new expression forms, \ourtool can still be 
 run on all programs: it just fails to improve annotations that require them.

\ourtool performs as follows on the Grift benchmarks:

\begin{itemize}

  \item \textbf{On 9 of 11 benchmarks}, \ourtool produces exactly the same
  type annotations as the hand-typed version.

  \item \textbf{N-body} defines a number of unused functions over vectors. Since
  they are under-constrained, \ourtool makes some arbitrary choices.
  On the reachable portion of the benchmark, we produce exactly the same
  type annotations as the hand-typed version.

  \item \textbf{Sieve} includes a library for stream processing, and the typed
  version of the benchmark gives streams a recursive type: $\mu~s.\kw{Int}
  \times (\rightarrow s)$. \ourtool migrates streams to type \tdyn, which
  forces the stream elements to have type \tdyn, due to ground constraints.

\end{itemize}

\subsection{Implementation and Performance}

The \ourtool{} tool is open-source and written in approximately 12,000 lines of
Rust. This code includes our new migration algorithm,  implementations of the
migration algorithms from \citet{rastogi:gti} and \citet{campora:migrating},
and a unified evaluation
framework that supports all the third-party tools that we use in our
evaluation. The evaluation framework is designed to automatically
validate the evaluation results we report. For example, to report that a migrated function is not
compatible with all untyped contexts, our framework requires an example of a context
that distinguishes between the migrated and original program, and runs both
programs in the given context to verify that they differ. The framework also
ensures that migrated programs are well-typed and structurally identical to
the original program.

We perform all our experiments on on a virtual machine with 4 CPUs and 8 GB
RAM, running on an AMD EPYC 7282 processor. The full suite consists of 892 LOC
and 33 programs. \ourtool{} produces migrations for our entire suite of
benchmarks in under three seconds.

\section{Related Work}
\label{sec:related}

There is a growing body of work on automating gradual type migration
 and related issues. Our work is most closely related to the four 
 algorithms we evaluate in \cref{sec:evaluation}.
\Citet{siek:gti} substitutes metavariables that appear in
type annotations with concrete types, using a variation on unification.
\Citet{rastogi:gti} builds a type inference system for ActionScript. Their
system ensures that inference never fails and produces types that are 
compatible with all untyped contexts.
\Citet{campora:migrating} uses variational typing to heuristically tame 
the exponential search space of types~\cite{chen:variational}.
\Citet{migeed:decidable} present decidability results for several type 
migration problems, including finding maximally precise migrations.

The aforementioned work relies on custom constraint solving algorithms. A
key contribution of this paper is an approach to gradual type migration using
an off-the-shelf MaxSMT solver, which makes it easier to build a type migration
tool. In addition, we present a comprehensive evaluation comparing all five approaches. 
As part of this effort, we have produced new, open-source implementations 
of the algorithms presented in \citet{rastogi:gti} and \citet{campora:migrating}.

\Citet{henglein:coercions} introduces the theory of coercions that we
use; \citet{scheme-to-ml} present an efficient compiler from Scheme to
ML that inserts coercions when necessary. This work also uses a custom
constraint solver and a complex graph algorithm.
The latter defines a \emph{polymorphic safety} criterion, which is
related to our notion of a context-restricted type migration
(Definition~\ref{def:ctx-migration}).
Coercions are equivalent to casts~\cite{siek:coercion}; both are
what \citet{vitousek:reticulated} calls the \emph{guarded} approach to
runtime type enforcement. \Citeauthor{vitousek:reticulated} introduces
the \emph{transient} approach, which only checks ground types/type
tags at runtime rather than wrapping/proxying used in the guarded
semantics. The transient approach is more efficient on the ``mixed
programs'' we generate but
offers weaker guarantees and can mask errors~\cite{greenman:tags,greenman:spectrum}.
\ourtool is based on the guarded model; we could reframe \ourtool to match transient by  only coercing between ground types.
\ourtoolp infers types based on particular elimination forms---just
like transient. Additionally, \ourtoolc makes no assumptions about
program contexts, just like transient's ``open world
soundness''~\cite{vitousek:runtime}.


\Citet{garcia:principal-gradual-types} extend \citeauthor{siek:gti}'s
work to infer principal types. Since we focus on monomorphic types, we
do not directly compare against their algorithm.
\Citet{miyazaki:dti} build on \citeauthor{garcia:principal-gradual-types}'s work, discussing the coherence 
issues what we point out in \cref{sec:what_matters}: types induce 
run-time checks that can affect program behavior. However, while 
we migrate all programs, \citeauthor{miyazaki:dti} use \emph{dynamic type
 inference} to discover type inconsistencies and report them as run-time errors.
\citet{castagna:perspective} propose another account of
gradual type inference that supports many features (let-polymorphism,
recursion, and set-theoretic types). They do not consider run-time safety.
Finally, \citet{campora:herder} extend their previous
work~\citeyearpar{campora:migrating} with a cost model for selecting migrations.
Like us, they discuss trade-offs in type migration, although they focus on type 
precision and performance, rather than semantics preservation.

\Citet{thf:typedscheme}, \citet{guha:flowtypes}, \citet{chugh:systemd}, and \citet{vekris:two-phase-typing}
are examples of retrofitted type checkers for untyped languages that feature
flow-sensitivity. These tools require programmers to manually migrate
their code, while we focus on automatic type migration. However, they go 
beyond our work by considering flow-sensitivity. 

\Citet{anderson:inference} present type inference for a representative 
fragment of JavaScript. However, the approach is not designed for gradual 
typing, where portions of the program may be untyped. Similarly, \citet{chandra:static-js} 
infer types for JavaScript programs with the goal of compiling them to run 
efficiently on low-powered devices; their approach is not gradual by design
and deliberately rejects certain programs.

\Citet{wies:errors} formulate a MaxSMT problem to localize OCaml type errors.
We also use MaxSMT and encode types in a similar manner. However, both the 
form of our constraints and the role of the MaxSMT solver are very different. 
In error localization, the MaxSMT problem helps isolate type errors from
well-typed portions of the program. In our work, the entire program must be
well-typed. Moreover, our constraints allow several typings, and we use soft
constraints to guide the MaxSMT solver towards solutions with fewer coercions.

Soft Scheme~\cite{wright:soft-typing} infers types for Scheme programs.
However, its type system is significantly different from the GTLC, which
hinders comparisons to contemporary type migration tools for the GTLC. 
\Citet[p. 41]{flanagan-thesis}'s discussion of how Soft Scheme's
sophistication can lead to un-intuitive types inspired work on set-based 
analysis of Scheme programs: \citet{flanagan:spidey} map program points to
 sets of abstract values, rather than types. 

Thorn~\cite{bard:thorn} presents an approach to gradual typing where
ordinary typed expressions cannot have runtime type errors, thus do not require
runtime checks. Instead, the programmer
must use \emph{like types} at the interface between typed and untyped code,
where runtime type errors may occur. The GTLC does not make this
distinction manifest, but it is essential for a safe type migration: 
types introduced by a tool must not introduce new runtime failures.

\ourtool{} relies on a traditional approach to constraint generation: we
carefully write constraint generation rules by hand. It is possible to
complement hand-written rules with additional sources of information to get
significantly better results. Some work uses run-time profiling to guide type
inference \cite{furr:drubyoopsla,rubydust,saftoiu:jstrace}. More recent work
uses programmer-supplied heuristics to guide type inference to produce more
readable results~\cite{red:inferdl,ren:hummingbird}. These approaches preserve
type soundness. It is also possible to produce type migrations using supervised
machine
learning~\cite{hellendoorn:dlti,nl2type,pradel:typewriter,wei:lambdanet}.

\section{Conclusion}
\label{sec:conclusion}

We present \ourtool{}, a new approach to type migration for the GTLC that is 
more flexible than previous approaches in two key ways. First, we formulate 
constraints for an off-the-shelf MaxSMT solver rather than building a
 custom constraint solver, which makes it easier to extend \ourtool{}. We 
 demonstrate this flexibility by adding support for several language features 
 beyond the core GTLC. Second, \ourtool{} can produce alternative migrations
 that prioritize different goals, such as type precision and compatibility
 with unmigrated code. This makes \ourtool{} a more flexible approach,
 suitable for migration in multiple contexts.
 
 We also contribute to the evaluation of type migration algorithms. We
 define a multi-stage evaluation process that accounts for multiple goals
 of type migration. We present a ``type migration challenge set'': a benchmark
 suite designed to illustrate the strengths and weaknesses of various type
 migration algorithms. We evaluate \ourtool{} alongside four existing type
 migration systems. Toward this end, we contribute open-source implementations
 of two existing algorithms from the literature, which we incorporate into a
 unified framework for automated type migration evaluation. We hope these
 evaluation metrics, new benchmarks, and benchmarking framework will aid future
 work by illuminating the differences among the many approaches to gradual
 type migration.

 \section*{Acknowledgements}

 We thank the OOPSLA reviews for their thoughtful feedback. We thank Aseem
 Rastogi and Zeina Migeed for helpful discussions about their work. We thank
 Matthias Felleisen and Shriram Krishnamurthi for reading early drafts, and
 discussing their experience with Soft Scheme and MrSpidey. We thank Laurence
 Tratt for help with \texttt{grmtools}~\cite{grmtools}, which
 \ourtool{} uses significantly. This work is partially supported by the National
 Science Foundation under grants CCF-2102288 and CCF-2129344.

\bibliography{bibexport}

\ifarxiv

\appendix

\section{Constraint Generation for Additional Expressions}\label{cg_appendix}

{
\figsize
\(
\begin{array}{c}
\inferrule*[Left=Sequence]
{\Gamma \vdash e_1 \Rightarrow T_1,\phi_1\\ \Gamma \vdash e_2 \Rightarrow T_2, \phi_2}
{\Gamma\vdash e_1; e_2 \Rightarrow e_1'; e_2', T_2, \phi_1 \wedge \phi_2}
\\[2em]

\inferrule*[Left=Let]
{\Gamma \vdash e_1 \Rightarrow T_1,\phi_1\\ 
\Gamma,x:T_1 \vdash e_2 \Rightarrow T_2, \phi_2}
{\Gamma\vdash \kw{let}~x = e_1~\kw{in}~e_2 \Rightarrow \kw{let}~x = e_1'~\kw{in}~e_2', T_2, \phi_1 \wedge \phi_2} \\[2em]

\inferrule*[Left=Fix]
{\Gamma \vdash e \Rightarrow T_1,\phi_1\\
 \textrm{$w$ and $\alpha$ are fresh} \\\\
 \phi_2 = (T_1 = \alpha \wedge w) \vee (T_1 = \tdyn \wedge \alpha = \tdyn \rightarrow \tdyn \wedge \neg w)}
{\Gamma\vdash \kw{fix}~f:\alpha . e \Rightarrow \kw{fix}~f:\alpha . [\escoerce(T_1,\alpha)]e', \alpha, \phi_1 \wedge \phi_2} \\[2em]

\inferrule*[Left=Pair]
{\Gamma\vdash e_1 \Rightarrow T_1, \phi_1\\
\Gamma\vdash e_2 \Rightarrow T_2, \phi_2 \\
\textrm{$\alpha$ and $w$ are fresh} \\\\
\phi_3 = (\alpha = \tpair{T_1,T_2} \wedge w) \vee (\alpha = \tdyn \wedge \eisground(\tpair{T_1,T_2}) \wedge \neg w)}
{\Gamma\vdash \tpair{e_1,e_2} \Rightarrow [\escoerce(\tpair{T_1,T_2},\alpha)]\tpair{e_1',e_2'}, \alpha, \phi_1 \wedge \phi_2 \wedge \phi_3}\\[2em]

\inferrule*[Left=First]
{\Gamma\vdash e \Rightarrow T,\phi_1 \\
\textrm{$\alpha$, $\beta$, and $w$ are fresh} \\\\
\phi_2 = ((T = \tpair{\alpha,\beta} \wedge w) \vee (T = \alpha = \tdyn \wedge \neg w))}
{\Gamma\vdash \kw{first}(e) \Rightarrow \kw{first}([\escoerce(T,\tpair{\alpha,\beta})]e'), \alpha,\phi_1 \wedge \phi_2}\\[2em]

\inferrule*[Left=Second]
{\Gamma\vdash e \Rightarrow T,\phi_1 \\
\textrm{$\alpha$, $\beta$, and $w$ are fresh} \\\\
\phi_2 = ((T = \tpair{\alpha,\beta} \wedge w) \vee (T = \beta = \tdyn \wedge \neg w))}
{\Gamma\vdash \kw{second}(e) \Rightarrow \kw{second}([\escoerce(T,\tpair{\alpha,\beta})]e'), \beta,\phi_1 \wedge \phi_2}
\end{array}
\)
}

{\figsize
\(
\begin{array}{c}
\inferrule*[Left=Vector]
{\Gamma\vdash e_1 \Rightarrow T_1, \phi_1\\
\Gamma\vdash e_2 \Rightarrow T_2, \phi_2 \\
\textrm{$\alpha$, $w_1$, and $w_2$ are fresh} \\\\
\phi_3 = (\alpha = \tvec{T_1} \wedge w_1) \vee (\alpha = \tdyn \wedge ground(\kw{vector}(T_1) \wedge \neg w_1)\\\\
\phi_4 = (T_2 = \tint \wedge w_2) \vee (T_2 = \tdyn \wedge \neg w_2)}
{\Gamma\vdash \tvec{e_1, e_2}\Rightarrow [\escoerce(\tvec{T_1},\alpha)]\tvec{e_1', [\escoerce(T_2,\tint)]e_2'}, \alpha, \phi_1 \wedge \phi_2 \wedge \phi_3 \wedge \phi_4}\\[2em]

\inferrule*[Left=VecGet]
{\Gamma\vdash e_1 \Rightarrow T_1, \phi_1\\
\Gamma\vdash e_2 \Rightarrow T_2, \phi_2 \\
 \textrm{$\alpha$, $w_1$ and $w_2$ are fresh} \\\\
 \phi_3 = (T_2 = \tint \wedge w_1) \vee (T_2 = \tdyn \wedge \neg w_1) \quad
 \phi_4 = (T_1 = \alpha = \tdyn \wedge \neg w_2) \vee (T_1 = \kw{vector}(\alpha) \wedge w_2)}
{\Gamma\vdash \kw{VecGet}(e_1, e_2)\Rightarrow \kw{VecGet}([\escoerce(T_1,\kw{vector}(\alpha))]e_1', [coerce(T_2,\tint)]e_2'), \alpha,\phi_1 \wedge \phi_2 \wedge \phi_3 \wedge \phi_4}\\[2em]

\inferrule*[Left=VecSet]
{\Gamma\vdash e_1 \Rightarrow T_1, \phi_1\\
\Gamma\vdash e_2 \Rightarrow T_2, \phi_2\\
\Gamma\vdash e_3 \Rightarrow T_3, \phi_3\\ 
 \textrm{$\alpha$, $w_1$ and $w_2$ are fresh} \\\\
 \phi_4 = (T_2 = \tint \wedge w_1) \vee (T_2 = \tdyn \wedge \neg w_1) \\
\phi_5 = (T_1=\kw{vector}(\alpha) \wedge T_2 = \alpha \wedge w_2) \vee 
(\alpha = \tdyn \wedge \eisground(T_1) \wedge \eisground(T_2) \wedge \neg w_2)}
{\Gamma\vdash \kw{VecSet}(e_1, e_2, e_3)\Rightarrow \kw{VecSet}([\escoerce(T_1,\kw{vector}(\alpha))]e_1', [\escoerce(T_2,\alpha)]e_2'), [\escoerce(T_3,\tint)]e_3'), \\\kw{vector}(\alpha),\phi_1 \wedge \phi_2 \wedge \phi_3 \wedge \phi_4 \wedge \phi_5}\\[2em]

\inferrule*[Left=Length]
{\Gamma\vdash e_1 \Rightarrow T_1, \phi_1\\
\Gamma\vdash e_2 \Rightarrow T_2, \phi_2 \\
 \textrm{$\alpha$, $w_1$ and $w_2$ are fresh} \\\\
 \phi_3 = (T_2 = \tint \wedge w_1) \vee (T_2 = \tdyn \wedge \neg w_1) \quad
\phi_4 = (T_1 = \kw{vector}(\alpha) \wedge w_2) \vee (\alpha = \tdyn \wedge \neg w_2)}
{\Gamma\vdash \kw{Length}(e_1, e_2)\Rightarrow \kw{Length}([\escoerce(T_1,\kw{vector}(\alpha))]e_1', [\escoerce(T_2,\tint)]e_2'), \tint,\phi_1 \wedge \phi_2 \wedge \phi_3 \wedge \phi_4)}
\end{array}
\)
}

\section{Typing the Language with Explicit Coercions}

The rules below define type-checking for the intermediate language of GTLC,
where all coercions are explicit.

\(
\begin{array}{@{\quad\quad\quad\quad}l@{\quad\quad\quad\quad\quad\quad}l}
\fbox{$\Gamma \vdash e : T$} \\[1em]
\inferrule*[Left=T-Id]{\Gamma(x) = T}{\Gamma \vdash x : T} 
&
\inferrule*[Left=T-Const]{\phantom{.}}{\Gamma \vdash c : \mathit{ty}(c)}
\\[1em]
\inferrule*[Left=T-Fun]{\Gamma,x:S \vdash e : T}{\Gamma \vdash \efun{x}{S}{e} : S \rightarrow T}
&
\inferrule*[Left=T-App]{
\Gamma \vdash e_1 : T_1 \rightarrow T_2 \\
\Gamma \vdash e_2 : T_1 }
{\Gamma\vdash e_1 e_2 : T_{2}}
\\[1em]
\inferrule*[Left=T-Mul]{
\Gamma \vdash e_1 : \tint \\
\Gamma \vdash e_2 : \tint
}
{\Gamma \vdash e_1 \times e_2 : \tint
}
&
\inferrule*[Left=T-Coerce]{
\vdash k : T_1 \rightarrow T_2 \\
\Gamma \vdash e : T_1 \quad
}
{\Gamma\vdash [k] e : T_{2}} \\[1em]
\fbox{$\vdash k : T$} \\[1em]
\inferrule*[Left=TC-Tag-Fun]{\phantom{.}}{\vdash \kw{fun!} : (\tdyn\rightarrow\tdyn) \rightarrow \tdyn}
&
\inferrule*[Left=TC-Chk-Fun]{\phantom{.}}{\vdash \kw{fun?} : \tdyn \rightarrow (\tdyn\rightarrow\tdyn)}
\\[1em]
\inferrule*[Left=TC-Tag-Int]{\phantom{.}}{\vdash \kw{int?} : \tdyn \rightarrow \tint}
&
\inferrule*[Left=TC-Chk-Int]{\phantom{.}}{\vdash \kw{int!} : \tint \rightarrow \tdyn}
\\[1em]
\inferrule*[Left=TC-Seq]{\vdash k_1 : T_1 \rightarrow T_2 \\ \vdash k_2 : T_2 \rightarrow T_3}{\vdash k_1;k_2 : T_1 \rightarrow T_3}
&
\inferrule*[Left=TC-Id]{\phantom{.}}{\vdash \kw{id}_T : T \rightarrow T} \\[1em]
\multicolumn{2}{c}{\inferrule*[Left=TC-Wrap]{\vdash k_1 : T_1 \rightarrow S_1 \\ \vdash k_2 : S_2 \rightarrow T_2 }
{\vdash \kw{wrap}(k_1, k_2) : (S_1 \rightarrow S_2) \rightarrow (T_1 \rightarrow T_2)}}
\end{array}
\)

\section{Soundness of found models}

Here we prove that if coercion insertion has a satisfiable model, it
induces a well typed coercion term.

%

\begin{lemma}[Coercions are well typed]
\label{lem:coercions-well-typed}
$\vdash \textrm{coerce}(S, T) : S \rightarrow T$ using the definition
of $\textrm{coerce}$ from Figure~\ref{baby-coercion-insertion}.
\begin{proof}
By induction on the sum of the sizes of the two coercions, with cases
drawn from the function. Let $\tdyn$ have size 1 and \tint have size $2$.
\begin{trivlist}
\item[($S=T$)] We have $\vdash \kw{id}_S : S \rightarrow T$ by \textsc{TC-Id}.
\item[($S=\tdyn$, $T=\tint$)] We have $\vdash \tint? : \tdyn \rightarrow \tint$ by \textsc{TC-Chk-Int}.
\item[($S=\tint$, $T=\tdyn$)] We have $\vdash \tint! : \tint \rightarrow \tdyn$ by \textsc{TC-Tag-Int}.
\item[($S=\tdyn$, $T=\tdyn\rightarrow\tdyn$)] We have $\vdash \kw{fun}? : \tdyn \rightarrow (\tdyn \rightarrow \tdyn)$ by \textsc{TC-Chk-Fun}.
\item[($S=\tdyn\rightarrow\tdyn$, $T=\tdyn$)]  We have $\vdash \kw{fun}! : (\tdyn \rightarrow \tdyn) \rightarrow \tdyn$ by \textsc{TC-Tag-Fun}.
\item[($S=S_1\rightarrow S_2$, $T=T_1\rightarrow T_2$)]
By the IH on $T_1$ and $S_1$, we have $\vdash \textrm{coerce}(T_1,S_1) : T_1 \rightarrow S_1$;
by the IH on $S_2$ and $T_2$, we have $\vdash \textrm{coerce}(S_2,T_2) : S_2 \rightarrow T_2$.
By \textsc{TC-Wrap} on these coercions, we have
  $\vdash \kw{wrap}(\textrm{coerce}(T_1,S_1),\textrm{coerce}(S_2,T_2))
  : (S_1 \rightarrow S_2) \rightarrow (T_1 \rightarrow T_2)$.
\item[($S=\tdyn$, $T=T_1\rightarrow T_2$)]
  By \textsc{TC-Chk-Fun}, we have $\vdash \kw{fun}?
  : \tdyn \rightarrow (\tdyn \rightarrow \tdyn)$.
  By the IH on $T_1$ and $\tdyn$ (which are smaller in total than our
  original function type and $\tdyn$), we know
  $\vdash \textrm{coerce}(T_1,\tdyn) : T_1 \rightarrow \tdyn$.
  Similarly, the IH on $\tdyn$ and $T_2$, we know
  $\vdash \textrm{coerce}(\tdyn,T_2) : \tdyn \rightarrow T_2$.
  By \textsc{TC-Wrap}, we have
  $\vdash \kw{wrap}(\textrm{coerce}(T_1,\tdyn),\textrm{coerce}(\tdyn,T_2)
  : (\tdyn \rightarrow \tdyn) \rightarrow (T_1 \rightarrow T_2)$.
  Finally, by \textsc{TC-Seq}, we can combine our first coercion with
  this to have
  $\vdash \kw{fun}?;\kw{wrap}(\textrm{coerce}(T_1,\tdyn),\textrm{coerce}(\tdyn,T_2))
  : \tdyn \rightarrow (T_1 \rightarrow T_2)$.
\item[($S=T_1\rightarrow T_2$, $T=\tdyn$)]
  By the IH on $\tdyn$ and $T_1$ (which are smaller in total than our
  original function type and $\tdyn$), we have
  $\vdash \textrm{coerce}(\tdyn, T_1) : \tdyn \rightarrow T_1$.
  Similarly, by the IH on $T_2$ and $\tdyn$, we have
    $\vdash \textrm{coerce}(T_2, ]tdyn) : T_2 \rightarrow \tdyn$.
  By \textsc{TC-Wrap}, we have
  $\vdash \kw{wrap}(\textrm{coerce}(\tdyn,T_1),\textrm{coerce}(T_2,\tdyn)) : (T_1 \rightarrow T_2) \rightarrow (\tdyn \rightarrow \tdyn)$.
  By \textsc{TC-Tag-Fun}, we have $\vdash \kw{fun}! :
  (\tdyn \rightarrow \tdyn) \rightarrow \tdyn$.
  Finally, we tie everything together with \textsc{TC-Seq}:
  $\vdash \kw{wrap}(\textrm{coerce}(\tdyn,T_1),\textrm{coerce}(T_2,\tdyn));\kw{fun}!
  : (T_1 \rightarrow T_2) \rightarrow \tdyn$.
  
\item[(otherwise)]
  If none of the other cases apply, we generate a coercion through
  $\tdyn$; such a coercion is doomed to fail. It is nevertheless well
  typed.
  First, observe that neither $S$ nor $T$ can be $\tdyn$, since one of
  the cases above would have adhered.
  So we can use the IH on $S$ and $\tdyn$ or $\tdyn$ and $T$, since
  every other type is larger than $\tdyn$.

  By the IH on $S$ and $\tdyn$, we have
  $\vdash \textrm{coerce}(S,\tdyn) : S \rightarrow \tdyn$.
  Similarly, by the IH on $\tdyn$ and $T$, we have
  $\vdash \textrm{coerce}(\tdyn,T) : \tdyn \rightarrow T$.
  By \textsc{TC-Seq}, we have
  $\vdash \textrm{coerce}(S,\tdyn);\textrm{coerce}(\tdyn,T) :
  S \rightarrow T$.
\qedhere

\end{trivlist}
\end{proof}
\end{lemma}

To keep things relatively neat notationally, we write $\sigma(X)$ to
mean applying $\textsc{Subst}(\sigma, -)$ to every indeterminate part of the
structure $X$, where $X$ might be a context $\Gamma$, expression $e$,
or type $T$.

\begin{theorem}[Models produce well typed terms (\Cref{thm:soundness})]
  \label{thm:pf:models-wf}
If $\Gamma \vdash e \Rightarrow e', T, \phi$ and $\sigma$ is a model
for $\phi$, then $\sigma(\Gamma) \vdash \sigma(e') : \sigma(T)$.
\begin{proof}
  By induction on the coercion insertion judgment.
\begin{trivlist}
\item[(\textsc{Id})] By \textsc{T-Id}.

\item[(\textsc{Const})] By \textsc{T-Const}, \textsc{T-Coerce}, and Lemma~\ref{lem:coercions-well-typed}.

\item[(\textsc{Fun})] Since $\sigma$ is a model of $\phi_1 \wedge
  \phi_2$, it is also a model for $\phi_1$.
So by the IH on $e$, we have $\sigma(\Gamma), x:\sigma(\alpha) \vdash e : \sigma(T)$.
By Lemma~\ref{lem:coercions-well-typed}, we know that $\vdash \textrm{coerce}(\sigma(T), \sigma(\beta)) : \sigma(T) \rightarrow \sigma(\beta)$.
By \textsc{T-Coerce}, we have $\sigma(\Gamma), x:\sigma(\alpha) \vdash
[\textrm{coerce}(\sigma(T), \sigma(\beta))] e : \sigma(\beta)$.
Finally, by \textsc{T-Fun}, we have $\sigma(\Gamma) \vdash
\efun{x}{\sigma(\alpha)}{[\textrm{coerce}(\sigma(T), \sigma(\beta))]
  e} : \sigma(\alpha \rightarrow \beta)$.
  The outer coercion is typed by \textsc{T-Coerce} and Lemma~\ref{lem:coercions-well-typed}.

\item[(\textsc{App})] Since $\sigma$ is a model of $\phi_1 \wedge
  \dots \wedge \phi_3 \wedge \phi_4 \wedge \phi_5$, it is also a model for each $\phi_i$.
  By the IHs, we have:
  \[
  \sigma(\Gamma) \vdash \sigma(e_1) : \sigma(T_1)
  \qquad \text{and} \qquad
  \sigma(\Gamma) \vdash \sigma(e_2) : \sigma(T_2).
  \]
  By Lemma~\ref{lem:coercions-well-typed}, we know that:
  \[
  \vdash \textrm{coerce}(\sigma(T_1), \sigma(\alpha \rightarrow
  \beta)) : \sigma(T_1) \rightarrow \sigma(\alpha \rightarrow \beta)
  \]
  and $\vdash \textrm{coerce}(\sigma(\beta), \sigma(\gamma)) : \sigma(\beta) \rightarrow \sigma(\gamma)$.
  We know that $\sigma(T_2) = \sigma(\alpha)$ by $\phi_4$, so
  by applying \textsc{T-Coerce} on the function and \textsc{T-App}, we have:
  \[ \sigma(\Gamma) \vdash ([\textrm{coerce}(\sigma(T_1), \sigma(\alpha \rightarrow
  \beta))] \sigma(e_1))\ \sigma(e_2)) : \sigma(\beta) \]
  We account for the outer coercion with \textsc{T-Coerce} and Lemma~\ref{lem:coercions-well-typed}.
  
\item[(\textsc{Mul})] Since $\sigma$ is a model for $\phi_1 \wedge
  \phi_2$, it is also a model for $\phi_1$ and $\phi_2$. By the IHs, we have:
  \[
  \sigma(\Gamma) \vdash \sigma(e_1) : \sigma(T_1)
  \qquad \text{and} \qquad
  \sigma(\Gamma) \vdash \sigma(e_2) : \sigma(T_2).
  \]
  By Lemma~\ref{lem:coercions-well-typed}, we have:
  \[
  \vdash \textrm{coerce}(\sigma(T_1),\tint) : \sigma(T_1) \rightarrow \tint
  \qquad \text{and} \qquad
  \vdash \textrm{coerce}(\sigma(T_2),\tint) : \sigma(T_2) \rightarrow \tint
  \]
  By applying \textsc{T-Coerce} twice and \textsc{T-Mul}, we have:
    \[ \sigma(\Gamma) \vdash ([\textrm{coerce}(\sigma(T_1), \tint)] \sigma(e_1)) \times ([\textrm{coerce}(\sigma(T_2),
      \tint)] \sigma(e_2)) : \tint \]
  The outer coercion is typed by \textsc{T-Coerce} and Lemma~\ref{lem:coercions-well-typed}.
\qedhere
\end{trivlist}
\end{proof}
\end{theorem}

\section{Existence of models}
\newcommand{\ok}{\mathrel{\mathsf{ok}}}

We show that models always exist for well scoped programs.

First, we borrow the ``well scoped'' relation from
\citet{matthews:multi}. We then show that a fully dynamic model always
exists for such well scoped programs, and that it is stable under weakening.
Let $V$ be a set of variables. We say a term $e$ is well scoped if
$\emptyset \vdash e \ok$.

\[
\begin{array}{@{\quad\quad\quad\quad}l@{\quad\quad\quad\quad\quad\quad}l}
\fbox{$V \vdash e \ok$} \\[1em]
\inferrule*[Left=WS-Id]{x \in V}{V \vdash x \ok} 
&
\inferrule*[Left=WS-Const]{\phantom{.}}{V \vdash c \ok}
\\[1em]
\inferrule*[Left=WS-Fun]{V \cup \{ x \} \vdash e \ok}{V \vdash \efun{x}{S}{e} \ok}
&
\inferrule*[Left=WS-App]{
V \vdash e_1 \ok \\
V \vdash e_2 \ok }
{V \vdash e_1 e_2 \ok}
\\[1em]
\inferrule*[Left=WS-Mul]{
V \vdash e_1 \ok \\
V \vdash e_2 \ok
}
{V \vdash e_1 \times e_2 \ok
}
&\end{array}
\]

\newcommand{\dynctx}{\ensuremath{\mathsf{dynctx}}}
Let $\dynctx(V)$ be defined as the context that maps every variable in
$V$ to $\tdyn$:
\[ \begin{array}{rcl}
  \dynctx(\emptyset)  &=& \cdot \\
  \dynctx(\Gamma,x:T) &=& \dynctx(\Gamma),x:\tdyn \\
\end{array} \]

\begin{theorem}[Well scoped terms have dynamic models]
  \label{thm:dynamic-models-exist}
  If $V \vdash e \ok$, then there exist $e'$, $T$, and $\phi$ such
  that for all dynamic models $\sigma$:
  \begin{enumerate}
  \item $\dynctx(V) \vdash e \Rightarrow e', T, \phi$,
  \item $\phi$ is satisfiable in $\sigma$, and
  \item $\textsc{Subst}(\sigma, T) = \tdyn$.
  \end{enumerate}
  \begin{proof}
    By induction on the derivation of $V \vdash e \ok$. We must take
    the right disjunct of every constraint except for two: the outer
    coercion on variables and applications could safely take either
    disjunct.
\begin{trivlist}
\item[(\textsc{WS-Id})] We have $x \in V$, so $x:\tdyn \in
  \dynctx(V)$. By \textsc{Id}; whether we pick the left or right
  disjunct, we have $\alpha = \tdyn = \dynctx(V)(x)$ (and so we will
  always find $\textsc{Subst}(\sigma, \alpha) = \tdyn$) and $\phi$
  is satisfiable in all dynamic models.
  
\item[(\textsc{WS-Const})] We have $T = \mathit{ty}(c)$ and $\phi =
  \kw{true}$ by \textsc{Const}.
  Pick $\alpha = \tdyn$; we have $\phi = \kw{true}$. The former is
  just $\tdyn$ under $\textsc{Subst}$.

\item[(\textsc{WS-Fun})] We know that $V \cup \{x\} \vdash e \ok$; by
  the IH, we have $\dynctx(V),x:\tdyn \vdash e \Rightarrow
  e',T,\phi_1$ such that $\phi_1$ is satisfiable. Since $\alpha =
  \tdyn$ and $\textsc{Subst}(\sigma, T) = \tdyn$, we know that
  $\textsc{Subst}(\sigma, \alpha \rightarrow T) = \tdyn
  \rightarrow \tdyn$, so we have $\eisground(\alpha \rightarrow T)$.
  Pick $\beta = \tdyn$. We already know $\phi_1$ is satisfiable, as is
  the right disjunct of $\phi_2$.
  We have $\textsc{Subst}(\sigma, \tdyn) = \tdyn$ immediately.
  
\item[(\textsc{WS-App})] We know that $V \vdash e_1 \ok$ and $V \vdash
  e_2 \ok$.
  By the IH on $e_1$, we have $\dynctx(V) \vdash e_1 \Rightarrow
  e_1',T_1,\phi_1$ such that $\phi_1$ is satisfiable in dynamic models and
  $\textsc{Subst}(\sigma, T_1) = \tdyn$.
  Similarly, the IH on $e_2$ finds $\dynctx(V) \vdash e_2 \Rightarrow
  e_2',T_1,\phi_1$ such that $\phi_2$ is satisfiable in dynamic models and
  $\textsc{Subst}(\sigma, T_2) = \tdyn$.

  Since $\phi_1$ and $\phi_2$ are both satisfiable in all models where
  all variables map to $\tdyn$, so is $\phi_1 \wedge \phi_2$.
  Pick $\alpha = \beta = \gamma = \tdyn$. We satisfy the right
  disjunction of $\phi_3$, and we've already established $\phi_4$
  (because $T_2$ will substitute to $\tdyn$, which is exactly equal to
  $\alpha$). We could take either disjunction if $\phi_5$---we already
  know $\beta = \tdyn$, so $\gamma = \tdyn$ either way.
  We have $\textsc{Subst}(\sigma, \gamma) = \tdyn$
  immediately.
  
\item[(\textsc{WS-Mul})] We know that $V \vdash e_1 \ok$ and $V \vdash
  e_2 \ok$.
  By the IH on $e_1$, we have $\dynctx(V) \vdash e_1 \Rightarrow
  e_1',T_1,\phi_1$ such that $\phi_1$ is satisfiable in dynamic models and
  $\textsc{Subst}(\sigma, T_1) = \tdyn$.
  Similarly, the IH on $e_2$ finds $\dynctx(V) \vdash e_2 \Rightarrow
  e_2',T_1,\phi_1$ such that $\phi_2$ is satisfiable in dynamic models and
  $\textsc{Subst}(\sigma, T_1) = \tdyn$.

  Since $\phi_1$ and $\phi_2$ are both satisfiable in all models where
  all variables map to $\tdyn$, so is $\phi_1 \wedge \phi_2$.
  Picking $\alpha = \tdyn$, we take the right disjuncts of $\phi_3$,
  $\phi_4$, and $\phi_5$. We have $\textsc{Subst}(\sigma, \alpha)
  = \tdyn$ immediately.
  \qedhere
\end{trivlist}
  \end{proof}
\end{theorem}

\begin{lemma}[Dynamic terms are stable under weakening]
  \label{lem:dynamic-models-stable} ~ \\
  If $\textsc{Subst}(\sigma, T) = \tdyn$, then
  $\textsc{Weaken}(\tdyn, T)$ is satisfiable.
  \begin{proof}
    Immediate: $\textsc{Weaken}(\tdyn, T) = P(\tdyn, T, \kw{true}) =
    \kw{true}$.
  \end{proof}
\end{lemma}

\begin{corollary}
  \label{cor:stable-models-exist}
  If $V \vdash e \ok$, then it has a satisfiable model that is stable
  under weakening.
  \begin{proof} The term $e$ has a dynamic model
    (Theorem~\ref{thm:dynamic-models-exist}) and dynamic models are
    stable under weakening
    (Lemma~\ref{lem:dynamic-models-stable}).
  \end{proof}
\end{corollary}

\fi

\end{document}